\documentclass[aps,groupedaddress,superscriptaddress,reprint,nofootinbib,amsmath,amsfonts,amssymb,prd,floatfix]{revtex4-2}

\usepackage{graphicx}
\usepackage{dcolumn}
\usepackage{bm}
\usepackage{xcolor}
\usepackage[mathlines]{lineno}
\usepackage{subfigure}
\usepackage[colorlinks=True,allcolors=blue]{hyperref}
\usepackage{tabularx}
\usepackage{xspace}

\begin{document}

\renewcommand*\thefootnote{\alph{footnote}}

\title{Detection prospects for multi-GeV neutrinos from collisionally heated GRBs}
\author{A. Zegarelli}
    \email{angela.zegarelli@roma1.infn.it}
 \affiliation{Dipartimento di Fisica, Università Sapienza, P.le Aldo Moro 2, I-00185 Roma, Italy,}
 \affiliation{INFN, Sezione di Roma, P.le Aldo Moro 2, I-00185 Roma, Italy}
\affiliation{Dipartimento di Fisica, Università di Roma “Tor Vergata”, Via della Ricerca Scientifica, 00133 Roma, Italy}

\author{S. Celli}%
 \affiliation{Dipartimento di Fisica, Università Sapienza, P.le Aldo Moro 2, I-00185 Roma, Italy,}
 \affiliation{INFN, Sezione di Roma, P.le Aldo Moro 2, I-00185 Roma, Italy}
 
 \author{A. Capone}%
 \affiliation{Dipartimento di Fisica, Università Sapienza, P.le Aldo Moro 2, I-00185 Roma, Italy,}
 \affiliation{INFN, Sezione di Roma, P.le Aldo Moro 2, I-00185 Roma, Italy}
 
   \author{S. Gagliardini}%
 \affiliation{Dipartimento di Fisica, Università Sapienza, P.le Aldo Moro 2, I-00185 Roma, Italy,}
 \affiliation{INFN, Sezione di Roma, P.le Aldo Moro 2, I-00185 Roma, Italy}
 
  \author{S. Campion}%
 \affiliation{Dipartimento di Fisica, Università Sapienza, P.le Aldo Moro 2, I-00185 Roma, Italy,}
 \affiliation{INFN, Sezione di Roma, P.le Aldo Moro 2, I-00185 Roma, Italy}
 
  \author{I. Di Palma}%
 \affiliation{Dipartimento di Fisica, Università Sapienza, P.le Aldo Moro 2, I-00185 Roma, Italy,}
 \affiliation{INFN, Sezione di Roma, P.le Aldo Moro 2, I-00185 Roma, Italy}

\date{\today}

\begin{abstract}
Neutrinos with energies ranging from GeV to sub-TeV are expected to be produced in Gamma-Ray Bursts (GRBs) as a result of the dissipation of the jet kinetic energy through nuclear collisions occurring around or below the photosphere, where the jet is still optically thick to high-energy radiation. So far, the neutrino emission from the ‘inelastic collisional model’ in GRBs has been poorly investigated from the experimental point of view. 
In the present work, we discuss prospects for identifying neutrinos produced in such collisionally heated GRBs with the large volume neutrino telescopes KM3NeT and IceCube, including their low-energy extensions, KM3NeT/ORCA and DeepCore, respectively. To this aim, we evaluate the detection sensitivity for neutrinos from both individual and stacked GRBs, exploring bulk Lorentz factor values ranging from 100 to 600. As a result of our analysis, individual searches appear feasible only for extreme sources, characterized by gamma-ray fluence values at the level of $F_{\gamma} \geq 10^{-2}~\rm erg~cm^{-2}$. In turn, it is possible to detect a significant flux of neutrinos from a stacking sample of $\sim$ 900 long GRBs (that could be detected by current gamma-ray satellites in about five years) already with DeepCore and KM3NeT/ORCA. The detection sensitivity increases with the inclusion of data from the high-energy telescopes, IceCube and KM3NeT/ARCA, respectively. 
\end{abstract}


\maketitle


\section{Introduction} \label{sec:introduction}
Gamma-Ray Bursts (GRBs) are the most luminous astrophysical phenomena currently observed in the Universe. They appear with a rate of the order of few per day, at random locations in the sky \cite{band,kouveliotou,meegan}. The released energy amounts up to $\sim$ 10$^{54}$ erg \cite{max_e_grbs,max_e_grbs_erratum} in a time lasting from a fraction of a second to several thousands of seconds. Since their serendipitous discovery in the late 1960s \cite{firstGRB}, gamma-ray satellites have been detecting the so-called GRB \emph{prompt}
radiation, a non-thermal high-energy emission characterized by a gamma-ray energy flux peaking at few hundreds keV in the observer frame and that occasionally extends in a long tail up to the GeV band \cite{GRBreview}. When broadband data are available, a typical GRB spectrum, $dN/dE$, can be fitted between $\sim$~10 keV and 10$^4$ keV by two smoothly connected power laws, also known as the Band function \cite{band}, with typical low and high-energy spectral slopes $\alpha\sim-1.0$ and $\beta\sim-2.5$, respectively.

The origin of the emission mechanism powering GRBs has been object of active debates since the early days of their discovery. The commonly accepted picture is the so-called \emph{fireball model} \cite{piran1999}, where a mildly-relativistic outflow is launched by the compact central engine, most likely a newly formed black hole \cite{GRBreview}. Substantial efforts have been made to model the jet outflow dynamics and the subsequent radiative processes, giving rise to a variety of scenarios, such as internal shocks \cite{is1,is2}, dissipative photospheres \cite{ph_model} and magnetic reconnection phenomena occurring above the photosphere \cite{mag_rec_1,mag_rec_2,mag_rec_3,mag_rec_4}.

Internal shock models are particularly suited for the description of the prompt emission: the powerful and collimated jet produced in the explosion converts a fraction of its kinetic energy into internal energy through shocks occurring among non-uniform shells characterized by different Lorentz factor values \cite{bustamante_nature}, at a typical distance of $10^{13}$~cm from the central engine \cite{guetta2004}. Part of the energy that gets dissipated at the shocks is expected to be converted into non-thermal particles, through the acceleration of hadrons and leptons, that hence radiate via synchrotron and Inverse Compton (IC). In spite of its ability to explain most of the high-energy properties of the prompt emission phase, including time variability and energetics, some of the observed GRB spectra appear in conflict with this scenario. Indeed, the synchrotron model predicts GRB spectra with $\alpha \sim -1.5$, while the fitted values are often significantly harder. E.g. long GRBs (LGRBs), characterized by a gamma-ray duration longer than 2 seconds, show on average $\alpha \sim -1.0$ \cite{alpha_long_1,alpha_long_2,alpha_long_3,alpha_long_4,alpha_long_5}. Some GRBs are even showing spectra beyond the so-called \emph{synchrotron line-of-death} (i.e. $\alpha \geq -2/3$) \cite{preece}, especially short GRBs (SGRBs) with observed gamma-ray duration shorter than 2 seconds \cite{alpha_short_1,alpha_short_2}. 

The inconsistencies between the synchrotron model and the observations have revived photospheric models, where the emission takes place closer to the photosphere of the jet. Actually, the photospheric emission is a natural consequence of the fireball model, the fireball being an optically thick expanding plasma made of particles and photons \cite{piran1999}. Photospheric emission from highly relativistic outflows was early considered as an explanation for the prompt emission of GRBs \cite{ph1,ph3}. However, these models did not have a significant impact for many years, as the observed GRB spectra appeared to be non-thermal since their very first observations. In photospheric models, the emission is expected to be constituted by a freely expanding radiation-dominated outflow with a thermal spectrum (Planck-like) peaked at $\sim$1 MeV in the observer frame. Thus, to explain the observed non-thermal emission, the photospheric component was proposed either to be reprocessed into a non-thermal emission arising from optically thin regions (e.g. \cite{meszaros}) or to be the result of projection effects. In the former case, several mechanisms have been suggested to operate below the photosphere, e.g. kinetic energy dissipation due to shocks \cite{rees,lazzati,ryde}, collisional processes \cite{belobodorov,vurm}, or magnetic energy dissipation due to field line reconnection \cite{ph2,spruit,giannios,zhang_yan}. In the geometrical interpretation, in turn, the observed emission would result in a superposition of spectra generated by photons emitted from a vast range of radii and angles which are detected simultaneously, rendering the inferred photosphere location angle dependent \cite{ph_angle_dep,peer}. Overall, it appears clear that radiation produced at this stage is an unavoidable component in GRB emission.


The first clear observation of a narrow thermal component in a GRB spectrum occurred in 2009, within the bright long GRB 090902B, detected by the Gamma-ray Burst Monitor (GBM) and Large Area Telescope (LAT) instruments on-board the \textit{Fermi} observatory \cite{grb090902b}. A time-resolved spectral analysis of such burst revealed an initial peaked component, with a spectral shape resembling the Planck function, interpreted as a clear sign of photospheric origin, followed by a later broadening of the spectrum described by a Band function with $\alpha=-0.6$ \cite{ryde}. This would suggest that the photospheric emission lasts during the whole burst duration, with the contribution of an additional component making the spectrum non-thermal. This picture has been corroborated by the discovery of other GRBs with a non-thermal spectrum overlapping the thermal one, e.g. GRB 100724B \cite{grb100724b} and GRB 110721A \cite{grb110721a}. 
Given the complexity of the emission observed from the prompt phase of GRBs, both in terms of spectral and temporal features, it is likely that different radiative stages occur. A key issue that still remains to be addresses is to which extent photospheric emission has to be complemented by additional processes and how to identify these different spectral components from observations.

In the present work, we investigate the so-called \textit{inelastic collisional model} \cite{belobodorov}, where a baryonic jet gets significantly heated as it propagates away from the central engine because of inelastic $pn$ collisions among a proton component and a slower neutron one, taking place in the sub-photospheric region of the jet. These collisions, besides being responsible for energy dissipation, are also expected to originate neutrinos in the multi-GeV energy range \cite{derishev,bahcall,meszaros_multiGeV_nu,asano,bartos,murase}. On the other hand, internal shock models predict the existence of TeV-PeV neutrinos as a result of $p\gamma$ interactions among shock accelerated protons and the radiation field located in the optically thin region of the jet \cite{waxman1997,rachen1998,piran1999,guetta2004,murase2006,zhang2013}. 
Hence, if neutrinos were revealed in coincidence with a GRB, they would allow to discriminate among the leptonic and hadronic nature of radiation; additionally, the measurement of their characteristic energy could be the key to identify the origin of the GRB prompt radiation (e.g. internal shocks vs photospheric emission). Thus, it is crucial to investigate neutrino emissions in a broad energy range.
To date, temporal and spatial associations among GRBs and neutrinos have only been searched for in the high-energy domain, mostly because the operating large volume neutrino telescopes IceCube \cite{icecube} and ANTARES \cite{antares} are designed to be mainly sensitive to TeV-PeV neutrinos. So far, these searches have not found any $\gamma-\nu$ association (refer to \cite{antares_first_grb_analysis,bright_grb_antares,my_analysis} for ANTARES analyses and \cite{icecube_analysis,icecube_grb2016} for the IceCube ones). 
Thus, after about fifty years from the discovery of GRBs, the lack of $\gamma-\nu$ associations still prevents us from undoubtedly establishing the mechanism responsible for the GRB prompt emission. 


Low-energy neutrinos produced in collisionally heated GRBs, i.e. those explained by the inelastic collisional photospheric model, might contribute to solve the puzzle. Some investigations in this direction have lead to the theoretical calculations of detection prospects of such neutrinos with IceCube \cite{icecube}, the South Pole neutrino observatory, and its low-energy extension, namely DeepCore \cite{deepcore}, sensitive to neutrinos with energies as low as $\rm E_{\nu}\sim$10 GeV. From a sample of bursts observed by BATSE \cite{meegan}, predictions estimated a non-negligible chance for detecting 10-100 GeV neutrinos in 5-10 years by using combined IceCube and DeepCore data \cite{bartos}. Other authors \cite{murase} showed that few neutrino-induced events can be detected by analyzing $\sim$1000-2000 GRBs stacked in a decade. Motivated by this, a first all-flavor search for transient emission of 1-100 GeV neutrinos was carried out using three years of data collected by the IceCube-DeepCore detectors. No significant emission was found in this sample, and upper limits on the expected volumetric rate of the transient neutrino sources were obtained, by assuming neutrino spectra consistent with the sub-photospheric emission \cite{icecube_analysis_lowE,icecube_analysis_lowE_ICRC}. However, it is worth noting that this analysis is time-dependent and it only refers to individual GRBs characterized by a duration up to approximately 600~s, a mean neutrino energy of 100~GeV, and a bolometric energy of 10$^{52}$~erg. Recently, an extension of this analysis has been presented \cite{icecube_analysis_lowE_ICRC}.

We investigate here for the first time about the possibility to reveal multi-GeV neutrinos from collisionally heated GRBs with the new generation neutrino telescope, KM3NeT \cite{km3net_loi}, currently under construction at two sites in the depth of the Mediterranean Sea. This detector will be sensitive to neutrinos down to few GeV energies thanks to the denser and compact array named KM3NeT/ORCA. We consider both individual and stacked searches on LGRBs and SGRBs, additionally exploring different bulk Lorentz factor values, ranging from low-luminous GRBs (i.e. those with $\Gamma \sim 100$) to high-luminous ones ($\Gamma \sim 600$).


The paper is structured as it follows. In Sec.~\ref{sec:model} we present the inelastic collisional model, detailing the gamma-ray production expected within such framework, as well as the spectral properties of the predicted neutrino fluxes. In Sec.~\ref{sec:neutrino_detectors} we describe the effective areas of neutrino detectors able to investigate the model predictions, both the currently operative ones and those under construction, focusing on the low-energy extensions for which we derived analytical parametrizations. In Sec.~\ref{sec:performances}, we discuss the neutrino signal and background characteristics, focusing on those parameters that are crucial for clearly assessing GRB-neutrino detections. 
Afterwards, in Sec.~\ref{sec:prospects_neutrino_detections}, we derive detector sensitivities with respect to both individual and stacking GRB-neutrino searches, the latter spanning over a GRB sample expected to be collected by the operative gamma-ray satellites in about five years. Finally, we discuss our results in Sec.~\ref{sec:conclusion}.

\section{Gamma-ray and neutrino production in the GRB inelastic collisional model} \label{sec:model}
The basic assumptions of the inelastic collisional model considered in this work \cite{derishev,review2008,belobodorov} are: (i) the presence of a dense, hot and neutron-rich central engine \cite{derishev,belobodorov2003}; (ii) a non-magnetized baryonic jet (see \cite{vurm} for an extension of the model including also a magnetized jet). These two requirements are not far from being realized since GRB jets are possibly produced by hydrodynamic processes taking place in the accretion disk around a black hole or a neutron star, and the dissociation of nuclei by gamma-ray photons in the inner regions of the disk could produce free neutrons \cite{GRBreview}. Initially, neutrons and protons accelerate as a single fluid because of frequent nuclear collisions, while at a later expansion stage the jet evolves into the two-fluid or \textit{compound state}: a slower neutron component with Lorentz factor $\Gamma_n$ is embedded in a faster proton flow with $\Gamma>\Gamma_n$. This compound flow develops when the timescale for $pn$ collisions becomes longer than the jet expansion time, at radius $R_n$ \cite{derishev,bahcall,meszaros_multiGeV_nu,belobodorov2003}. The jet becomes transparent to radiation at the photosphere ($R_{\rm ph}\sim(10-20)R_n$), where the thermal emission is effective, as modified by the sub-photospheric collisional process. Such a heating mechanism, that is realized in the region of the jet between $R_n$ and $R_{\rm ph}$, injects energy into electron-positron pairs via two branches occurring at comparable heating rates: (i) electrons are heated by Coulomb collisions with protons and consequently radiate; (ii) inelastic $pn$ collisions. As a result, nuclear and Coulomb collisions in GRB jets create a hot $e^{\pm}$ plasma, that radiates its energy producing an escaping radiation with a well-defined spectrum. In the following, we will focus on inelastic nuclear collisions, as this channel is responsible for neutrino production within the photosphere. 

\subsection{Inelastic nuclear collisions} 
The region between $R_{n}$ and $R_{\rm ph}$ is characterized by inelastic nuclear collisions between protons and neutrons, significantly affecting the jet dynamics (sub-photospheric collisional heating), namely
\begin{equation}
\left\{\begin{matrix}
p+n\rightarrow p+p+\pi^{-} & \\
p+n\rightarrow n+n+\pi^+ & \\
\end{matrix}\right.,
\end{equation}
as well as
\begin{equation}
\left\{\begin{matrix}
p+p\rightarrow p+n+\pi^{+} & \\ 
p+p\rightarrow p+p+\pi^0 & \\
n+n\rightarrow p+n+\pi^{-}
\end{matrix}\right..
\end{equation}
The rate of $pn$ collisions per unit volume is given by \cite{belobodorov}:
\begin{equation}
\dot{\rm N}=n n_n \Gamma_{\rm rel} \sigma c,
\end{equation} where $\sigma=3\times 10^{-26}~\rm cm^2$ \cite{GRBreview} is the nuclear cross section, $n$ and $n_n$ are respectively the proton and neutron number densities, $c$ is the speed of light in vacuum, and $\Gamma_{\rm rel}$ is the relative Lorentz factor of the neutron and proton component of the jet, i.e.
\begin{equation}
\label{eq:rel_lorentz_factor}
\Gamma_{\rm rel}=\frac{1}{2}\left(\frac{\Gamma}{\Gamma_{ n}}+\frac{\Gamma_{n}}{\Gamma}\right)\simeq \frac{\Gamma}{2\Gamma_{n}},
\end{equation}
with $\Gamma \gg \Gamma_n$.
Each collision between protons and neutrons dissipates a fraction of kinetic energy, and quasi-thermal nucleons are produced with $ \rm E_{N,\rm c}^{\rm th}\simeq k_p \Gamma_{\rm rel} m_p c^2$ in the comoving frame of the interacting flow\footnote{Neutrons that survive these collisions travel to larger distances before decaying, possibly affecting the afterglow radiation from GRBs
\cite{belobodorov_afterglow}.}. 
Here $\rm k_p \approx 0.5$ is the nucleon inelasticity (i.e. the ratio between the inelastic and the total interaction cross section) \cite{murase}, and $\rm m_p$ the proton mass. A comparable amount of energy converts to mildly relativistic pions. Charged pions immediately decay into muons (in 26 ns), in turn unstable towards the production of electron/positron pairs: 
\begin{equation}
\label{eq:charged_pion_decay}
\pi^{\pm} \rightarrow \mu^{\pm} + \nu_{\mu}/\bar{\nu}_{\mu} \rightarrow e^{\pm} + \nu_e / \bar{\nu}_e.
\end{equation}
In addition, from the neutral pion decay, high-energy gamma rays are produced, that quickly convert to $e^{\pm}$:
\begin{equation}
\pi^0 \rightarrow \gamma+\gamma \rightarrow e^{\pm}.
\end{equation}
Such $e^{\pm}$ pairs, together with the ones produced by the aforementioned Coulomb collisions, can either up-scatter the thermal photons produced at the jet launch site to higher energies (via IC) and/or radiate via synchrotron emission, modifying the radiation spectrum and thus introducing a non-thermal component. The photon spectrum emitted by a collisionally heated jet was first derived by \cite{belobodorov}, through accurate  Monte Carlo simulations of the radiative transfer in the expanding jet. The resulting GRB spectra were shown to peak at $\sim$1 MeV 
and to extend at higher energies with a photon index $\beta \sim -2.5$, well reproducing the prompt observations \cite{preece,band}.

\subsection{Neutrino production at the source} 
\label{sec:neutrinoproduction}
From hadronic collisions in the sub-photospheric region of GRB jets, neutrinos are also produced (see Eq.~\eqref{eq:charged_pion_decay}). Charged pions on average carry 2/3 of the energy transferred by the protons in hadronic nuclear $pn$ collisions. By considering that neutrinos take $\sim 3/4$ of $\pi^{\pm}$ energy,
we can evaluate the average fraction of pion energy that is given to neutrinos, $f_{\nu}$, as:
\begin{equation}
\label{eq:fract_en_neutrinos}
f_{\nu} \sim \frac{2}{3} \cdot \frac{3}{4}=\frac{1}{2}.
\end{equation}
Neutrinos therefore carry away a significant fraction of the energy $ \rm E_{\rm k,diss,s}$ dissipated in inelastic nuclear collisions, where the subscript $s$ refers to the source rest frame, which is related to the comoving one through the Lorentz boost $ \rm E_{\rm s}=\Gamma E_{\rm c}$. The corresponding energy of the neutrino burst (which does not suffer any adiabatic cooling) is 
\begin{equation}
{\rm E_{\nu, \rm s}}=f_{\nu} \rm E_{\rm k,diss,s}\simeq \frac{1}{2}E_{\rm k,diss,s}.
\end{equation}
Therefore, the energy channeled into radiation produced by the GRB jet is the remaining
\begin{equation}
\label{eq:egamma}
{\rm E_{\gamma,\rm s}}=f_{\rm ad}(1-f_{\nu}){\rm E_{\rm k,diss,s}}\simeq \frac{f_{\rm ad}}{2}\rm E_{\rm k,diss,s},
\end{equation}where $f_{\rm ad}<1$ describes the reduction in radiation energy due to adiabatic cooling in the expanding opaque jet below the photosphere.\\
Hence, the ratio among the neutrino and the radiation burst energies (or their isotropic equivalents) is given by
\begin{equation}
\label{eq:w}
w=\frac{\rm E_{\nu, \rm s}}{\rm E_{\gamma, \rm s}}=\frac{\rm E_{\nu, \rm s}^{\rm iso}}{\rm E_{\gamma, \rm s}^{\rm iso}}=\frac{1}{f_{\rm ad}}.
\end{equation}
Assuming that half of the energy is dissipated in the adiabatic expansion (i.e. $f_{\rm ad}=0.5$), we expect
\begin{equation}
\label{eq:enu}
\rm E_{\nu, \rm s}=2E_{\gamma, \rm s} \rightarrow E_{\nu, \rm s}^{\rm iso}=2E_{\gamma, \rm s}^{\rm iso}.
\end{equation}
By defining the ratio $\xi_{\rm N}$ among the energy dissipated in inelastic nuclear collisions and the gamma-ray energy produced in the GRB jet as
\begin{equation}
\label{eq:xi}
\xi_{\rm N} = \frac{E_{\rm k,diss,s}^{\rm iso}}{E_{\gamma,\rm s}^{\rm iso}},
\end{equation}
the benchmark scenario with $f_{\rm ad}=0.5$ would imply $\xi_{\rm N}=4$ (see Eq.~\eqref{eq:egamma}). \\
In the present work, we consider only muon neutrino (and the corresponding anti-neutrino) emissions, since their interactions in charged current with nucleons inside large volume neutrino telescopes are well identified, resulting in long muon tracks.
By considering the energy carried by $\nu_{\mu}$ and $\bar{\nu}_\mu$ only, Eq.~\eqref{eq:enu} can be written as:
\begin{equation}
\label{eq:nu_energy}
\rm E_{\nu_{\mu}+\bar{\nu}_{\mu},\rm s}^{\rm iso} \sim \frac{2}{3}E_{\gamma, \rm s}^{\rm iso}.
\end{equation}
Thus, the energy going in muon neutrinos is estimated to be $\sim$67\% of the gamma-ray energy. Such a linear scaling implies that the absolute gamma-ray energy and the model parameter $\xi_{\rm N}$ in Eq.~\eqref{eq:xi} are crucial, as they influence the neutrino spectral normalization \cite{murase}. 

The energy of the emitted neutrinos is a function of the Lorentz factor of the jet, $\Gamma$, and of the relative Lorentz factor of the proton and neutron components (given in Eq.~\eqref{eq:rel_lorentz_factor}), through \cite{belobodorov,asano,bartos,murase}:
\begin{equation}
\label{eq:peak_energy}
\rm E_{\nu} \approx 0.1 \Gamma~\Gamma_{\rm rel}~\mathrm{m_p c}^2 \rightarrow E_{\nu} \simeq 100~\rm GeV \left(\frac{\Gamma}{500}\right) \left(\frac{\Gamma_{\rm rel}}{2}\right).
\end{equation}
This relation implies an average expected neutrino energy of $\rm E_{\nu}\sim$~10-100 GeV for Lorentz factors $\Gamma \sim 100-1000$. Therefore, measuring the neutrino energy would provide a direct handle on the Lorentz factor of the jet, which is a key in resolving GRB dynamics.\\The neutrino spectra arising from sub-photospheric collisionally heated model are quasi-thermal, hence their shape is bell-like. The exact details of neutrino spectra have been obtained with detailed Monte Carlo simulations including cooling processes of secondary mesons and leptons (hadronic losses, radiative cooling and adiabatic expansion) by \cite{murase}. The resulting spectra are further discussed in Sec.~\ref{sec:prospects_neutrino_detections}.

\section{Current and future low-energy neutrino detectors} \label{sec:neutrino_detectors}
The search for multi-GeV neutrinos from GRBs with Cherenkov telescopes needs a compact arrays of 3D photomultiplier sensors, in order to detect the Cherenkov light induced by the propagation of the relativistic particles produced in neutrino interactions.

The IceCube observatory, operating at the South Pole, is complemented by DeepCore \cite{deepcore}, an array characterized by a higher concentration of digital optical modules (DOMs), optimized for the detection of neutrinos with energies down to 10 GeV. DeepCore is constituted by 15 strings located in a radius of 125 m at a depth from $\sim$2100 m to $\sim$2450 m in the ice. Eigth strings are very close to the bottom center of IceCube, with a DOM-to-DOM vertical spacing varying between 7 and 17~m. The DeepCore detector is operational since about 10 years. A low-energy in-fill extension to IceCube has also been proposed, named PINGU \cite{pingu}, that will be characterized by an effective mass of about 6~Mton for neutrino energies above few GeV. So far, no public effective area is available for PINGU, thus it will not be considered in the following estimations.

Currently, another low-energy neutrino detector is under construction in the Northern hemisphere, off the Mediterranean France coast at about 2450 m depth, namely the KM3NeT/ORCA neutrino telescope. Its optical modules are being arranged in the dense configuration required for detecting events with energies as low as few GeV. This range is three orders of magnitude lower that the typical energy scale probed by the high-energy detector KM3NeT/ARCA (currently under construction offshore Sicily, in Italy), designed for neutrino astroparticle physics studies \cite{km3net_loi}. At the time of writing, KM3NeT/ORCA is taking data with 10 strings, with an average horizontal spacing between strings of about 20~m and a vertical spacing between DOMs of about 9~m
\cite{km3net_loi,km3net_design,last_km3net_orca}. Once completed, KM3NeT/ORCA will consist of 115 strings, arranged in a circular footprint with a radius of about 115 m.

The primary goal of low-energy neutrino detectors is to unveil the intrinsic properties of neutrinos, as the mass hierarchy, by investigating neutrino oscillation studies in the atmospheric sector. Still, the low energy domain offers interesting possibilities for exploring astrophysical science cases, e.g. the collisional heating mechanism powering the GRB prompt emission presented above. Hence we proceed by investigating the KM3NeT/ORCA and DeepCore performances in the context of multi-GeV GRB analyses. The following part of the current section presents an analytical parametrization of the detector effective areas, while the next section details about expected background rate in each detector.

\begin{figure}[t!]
    \centering
    \includegraphics[width=\columnwidth]{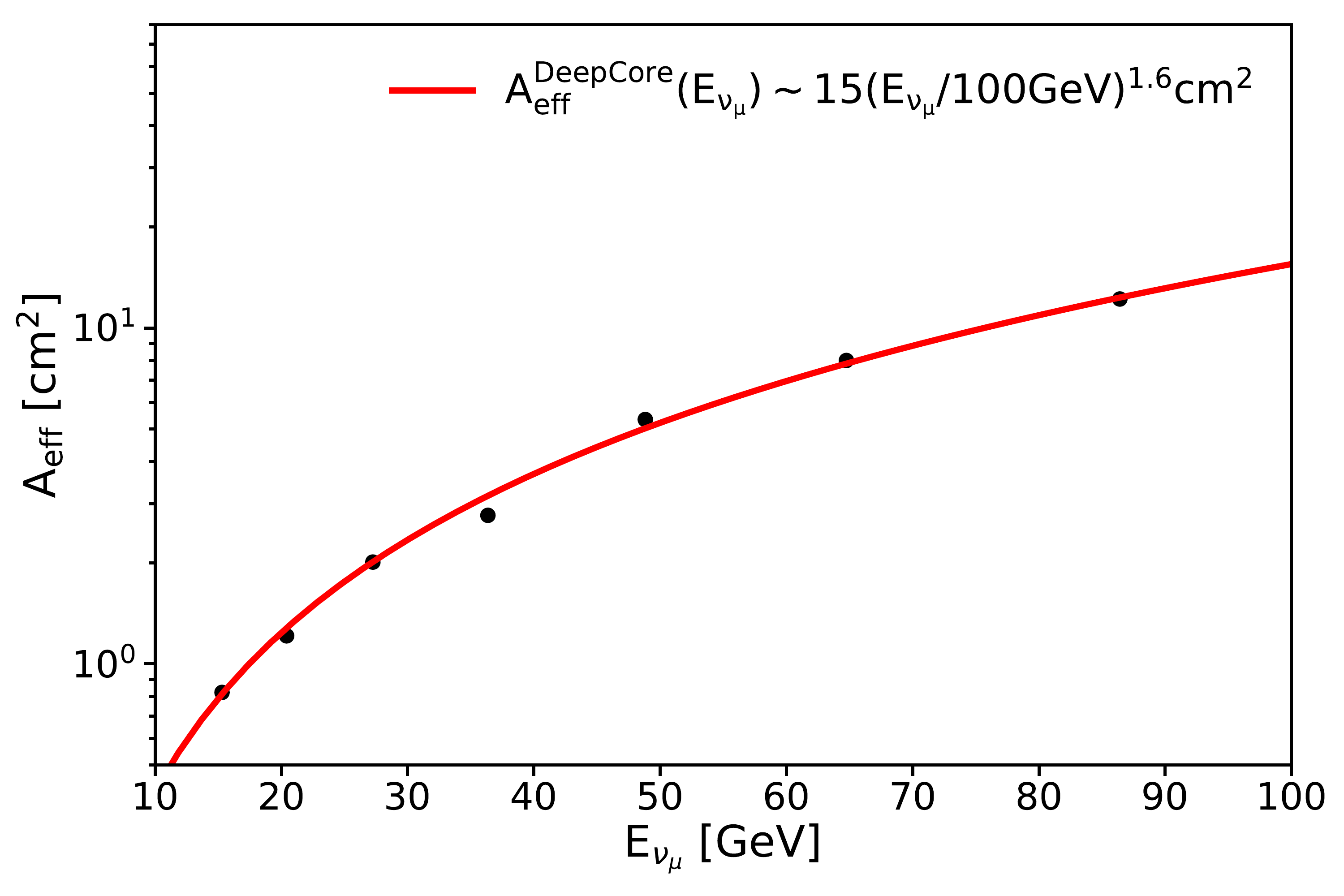}
    \caption{DeepCore effective area at trigger level for neutrino energies between $\sim$10 GeV and 100 GeV. Black points represent published values from the IceCube Collaboration \cite{deepcore}, while the red solid line shows the best fit obtained with Eq.~\eqref{eq:fit_deepcore}.}
    \label{fig:fit_deepcore}
\end{figure}
\begin{figure}[t!]
    \centering
    \includegraphics[width=0.95\columnwidth]{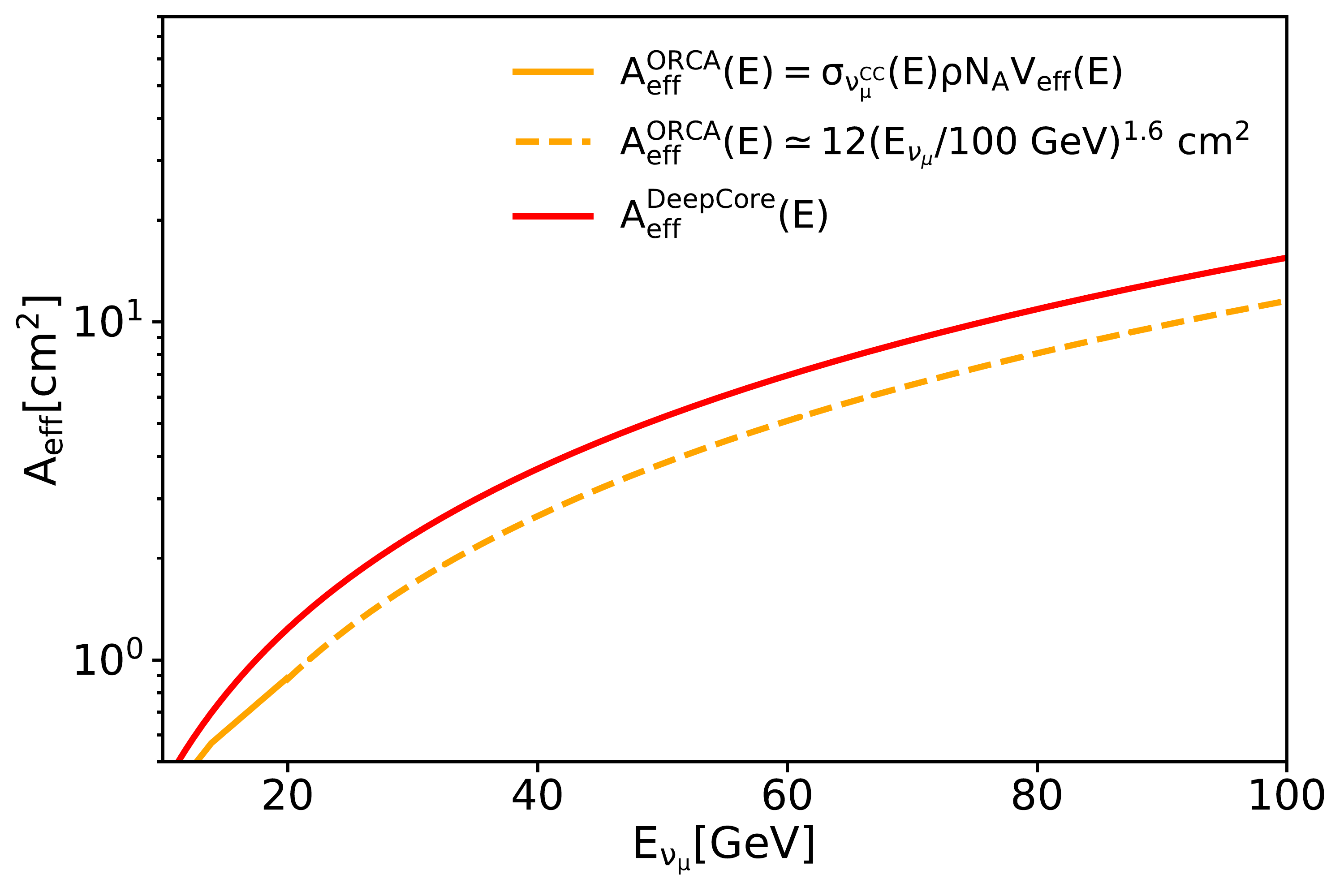}
    \caption{KM3NeT/ORCA effective area at trigger level for neutrino energies between $\sim$10 GeV and 100 GeV (orange line) as compared to the DeepCore effective area (red line). The orange solid line, extending up to $\sim$20 GeV, shows the computation resulting from Eq.~\eqref{eq:veff}, namely starting from the detector effective volume $\rm V_{eff}(E)$ \cite{new_veff_orca}. In turn, the orange dashed line $\rm E_{\nu_{\mu}}>20~GeV$ represents its extrapolation, by adopting the same energy dependence as in DeepCore.}
    \label{fig:aeff_orca}
\end{figure}

\subsection{Detector effective areas}
The present study is performed by considering instrument response functions of each detector at trigger level: in particular, effective areas for $\nu_{\mu}+\bar{\nu}_{\mu}$ events in IceCube-DeepCore and KM3NeT/ORCA-KM3NeT/ARCA are taken by \cite{deepcore} and \cite{km3net_loi}, respectively, accordingly with the actual detector configurations. Note that the KM3NeT/ORCA effective area at trigger level is not directly available from literature, but it can be obtained by knowing its effective volume $\rm V_{eff}$ towards muon neutrino events, which has been published for the complete detector configuration up to energies of $\sim 20$~GeV \cite{km3net_loi}. By defining the muon neutrino charged current cross section $\sigma_{\nu_{\mu}^{\rm CC}}(\rm E)$, the medium density $\rho$ (i.e. the water density at KM3NeT/ORCA site), and the Avogadro constant $\rm N_A$, we derived the KM3NeT/ORCA effective area as
\begin{equation}
\label{eq:veff}
\rm A_{\rm eff}(E)=\sigma_{\nu_{\mu}^{\rm CC}}(E) \rho N_A V_{\rm eff}(E) \, .
\end{equation}
Since our GRB-$\nu$ flux evaluation requires values up to $\sim$1 TeV, we further extrapolate the KM3NeT/ORCA effective area behaviour at higher energies than available, by using the same energy dependence as in DeepCore\footnote{This assumption can be considered valid as both detectors are characterized by a dense configuration of optical modules.}.
A best fit procedure to the DeepCore effective area results into:
\begin{equation}
\label{eq:fit_deepcore}
    \rm A_{eff}^{\rm DeepCore}(E_{\nu_{\mu}})=15\left(\frac{E_{\nu_{\mu}}}{100~\rm GeV}\right)^{1.6}~cm^2, 
\end{equation}
as shown in Fig.~\ref{fig:fit_deepcore}. Hence, we expect the KM3NeT/ORCA effective area at trigger level to correspond in the energy range $10-100$~GeV to (see Fig.~\ref{fig:aeff_orca})
\begin{equation}
\label{eq:fit_orca}
    \rm A_{eff}^{\rm ORCA}(E_{\nu_{\mu}})=12\left(\frac{E_{\nu_{\mu}}}{100~\rm GeV}\right)^{1.6}~cm^2.
\end{equation}
These parametrizations show that DeepCore and KM3NeT/ORCA are expected to have comparable performances. Note that performances at trigger level are the highest possibly achievable by experiments, later reduced by the efficiency of the event selection at analysis level. However, since none of the detectors under investigation has yet implemented analysis tailored at identifying low-energy neutrinos from cataloged GRBs, we will conservatively adopt the trigger level performances in the following, in the forms of Eqs.~\eqref{eq:fit_deepcore} and \eqref{eq:fit_orca}. Note also that triggering strategy are subject to change: e.g., recently, a new approach has been developed for KM3NeT/ORCA \cite{new_veff_orca,last_km3net_orca}, which is expected to increase the trigger efficiency in the few GeV neutrino energy range with respect to the case here considered.

\section{Signal and background estimation for GRB-neutrino detections} \label{sec:performances}
Quantitative estimations concerning detection prospects of low-energy neutrinos emerging from collisionally heated GRBs with current (DeepCore and IceCube) and under construction (KM3NeT/ORCA and KM3NeT/ARCA) neutrino telescopes are here presented.  As at multi-GeV energies the atmospheric background is severely limiting the identification of cosmic signals, in this search we will consider only upward going neutrinos. Indeed, Earth-filtered events allow to reduce significantly the atmospheric muon background. Additionally, we will consider only events due to $\nu_{\mu}$ charged current (CC) interactions: the muon originated in such interactions indeed results into a long track that allows to define with good accuracy the direction of the incoming neutrino. In turn, a worse directional reconstruction is expected for shower-like events, e.g. those originated by $\nu_e$ and $\nu_{\tau}$ CC interaction channels, as well as by all flavor neutral current (NC) interactions. The addition of this event topology in the present work would require to extend the search cone around each source (particularly for large values of $\Gamma$, see Sec.~\ref{sec:prospects_neutrino_detections}), implying a higher background level affecting the analysis, while at the same time allowing to probe the entirety of the neutrino flux reaching Earth. The exact balance among these two effects deserves a detailed investigation, that we defer to a future work because instrumental performances at such low energies are currently not available for all neutrino telescopes under exam.

For the purposes of this analysis, synthetic GRB characteristics are considered: in particular, as we aim at evaluating the neutrino flux expected on Earth from a GRB population powered by the collisional heating mechanism, we define a source sample reflecting the observed properties of the population.
Therefore, by collecting satellite's published data, we built distributions of several quantities, as the time interval over which a burst emits from 5\% of its total measured gamma-ray emission to 95\% (also known as $T_{90}$), and the gamma-ray fluence $F_\gamma$, that are key parameters for the background and signal estimation, respectively. Under the hypotheses of the inelastic collisional model (explained in Sec.~\ref{sec:model}), the neutrino signal would be produced in spatial and temporal coincidence with the prompt phase of GRBs. For this reason we conservatively define a temporal search window around each burst as wide as $T_{90} \pm 0.3T_{90}$

The neutrino spectra produced in collisionally heated GRBs are taken from \cite{murase} under the following assumptions: i) the neutrino and electromagnetic emission is released at the photosphere (i.e. at $R_{\rm ph}$), where the optical depth for the $pn$ reactions is close to unity; ii) the relative Lorentz factor between the proton and the neutron component is $\Gamma_{\rm rel}=3$ ($\Gamma\simeq6\Gamma_{n}$, see Eq.~\eqref{eq:rel_lorentz_factor});
iii) the fraction of gamma-ray energy dissipated in nuclear collision is $\xi_{\rm N}=4$. Note that $\Gamma_{\rm rel}$ influences the characteristic energy of emitted neutrinos, which scales as $\Gamma_{\rm rel}/2$ (Eq.~\eqref{eq:peak_energy}), namely a faster proton flow inside the jet would imply a higher value for the peak of neutrino spectra. In particular, with respect to the reference case with $\Gamma_{\rm rel}=3$, a proton flow with $\Gamma \simeq 20 \Gamma_{n}$ gives typical neutrino energy higher by a factor $\sim3$ once the jet Lorentz factor $\Gamma$ is fixed. In turn, $\xi_{\rm N}$, in Eq.~\eqref{eq:xi}, influences the neutrino fluence normalization, since it is related to the ratio among neutrino and gamma-ray energies, as demonstrated in Sec.~\ref{sec:neutrinoproduction}. The inelastic collisional model predicts at most $\xi_{\rm N} \approx 20$ \cite{murase}, which corresponds to $f_{\rm ad}\sim 0.1$ (see Eq.~\eqref{eq:egamma} and Eq.~\eqref{eq:xi}). In such a case, the neutrino fluence normalization would rise by a factor 5 with respect to the benchmark case here considered, thus increasing the expected GRB-emissivity rate of neutrinos. While the spectral details of emission, and hence the expected number of events in neutrino telescopes, might depend on the specific values of the model parameters assumed, the results of the neutrino sensitivity computation presented in the following are quite stable against reasonable variations of these parameters, i.e. $4\leq\xi_{\rm N}\leq20$ and $3\leq\Gamma_{\rm rel}\leq10$. In fact, the sensitivity is rather dominated by the large variety of the GRB population in terms of temporal and spectral properties (namely $T_{90}$ and $F_\gamma$, that affect respectively the number of background and signal events).

\begin{figure}[t!]
    \centering
    \includegraphics[width=\columnwidth]{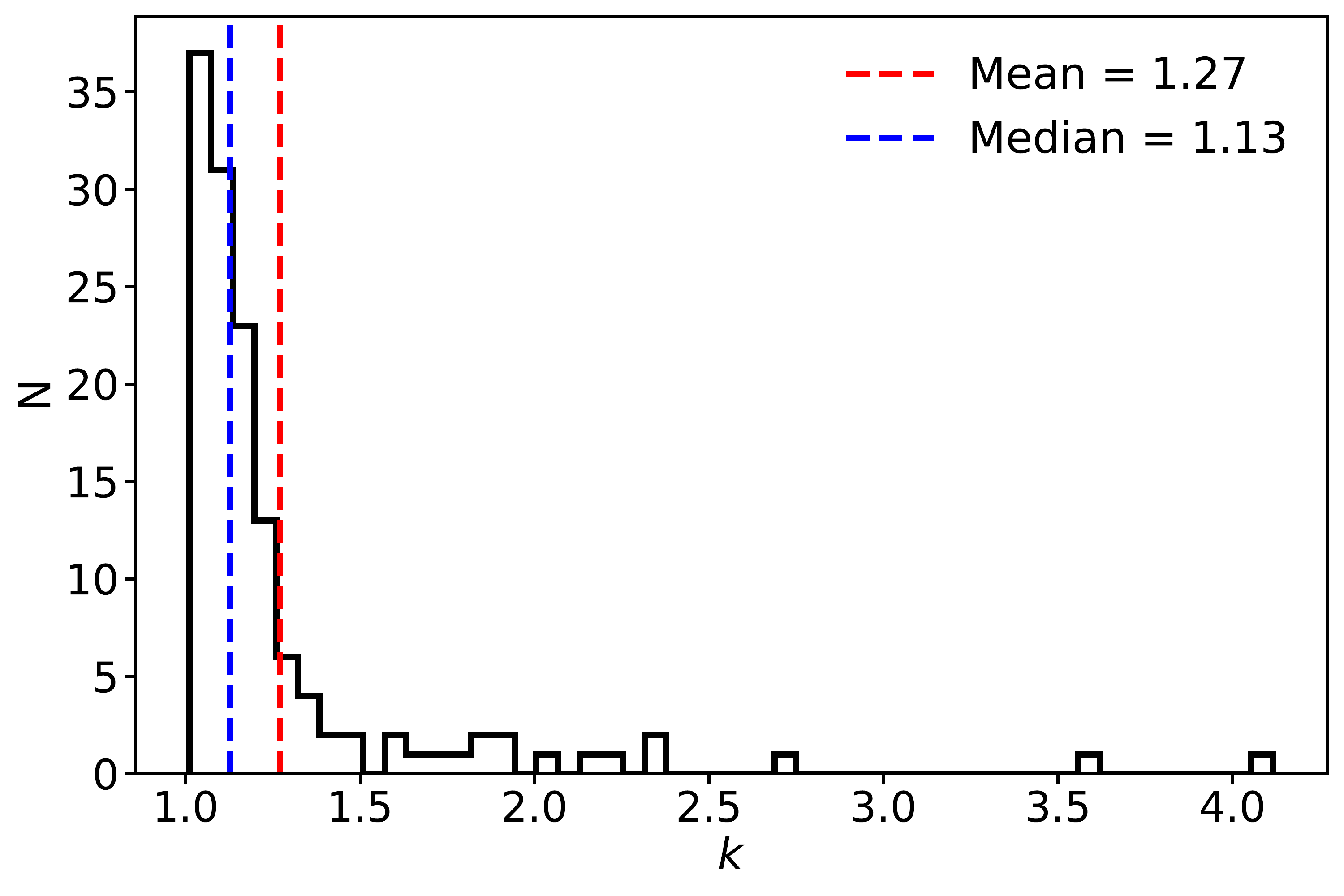}
    \caption{$k$-correction values calculated on the GRB sample collected by \textit{Fermi}-GBM during the first ten years of detector operation, including both LGRBs and SGRBs \cite{k_corrections}. The red and blue dashed lines show the mean and the median of the $k$-correction values, respectively.}
    \label{fig:k_corr}
\end{figure}

The observed neutrino fluence produced in collisionally heated GRBs from a source at redshift $z$ is 
\begin{equation}
\label{eq:nu_fluence_obs}
F_{\nu} \propto \frac{1+z}{4\pi d_L^2(z)} \int \rm dE \left({\rm E} \frac{\rm dN}{\rm dE}\right)_{\nu,\rm s} \, ,
\end{equation}
where $d_L(z)$ is the luminosity distance of the source and the factor $\frac{1+z}{4\pi d_L^2(z)}$ takes into account the cosmological distance of the source and the dilution of the neutrino energy flux from the source to Earth.
It is also worth noting that, as sub-photospheric gamma rays constitute the prompt emission, the neutrino fluence is proportional to the observed gamma-ray fluence through Eq.~\eqref{eq:nu_energy}, namely $F_\nu \sim F_{\gamma}$.
To define the GRB of the sample of our simulation, we extracted for each source the spectral parameters $F_{\gamma}$ and $T_{90}$ in accordance with \textit{Fermi}-GBM distributions\footnote{\textit{Fermi}-GBM catalogue: \url{https://heasarc.gsfc.nasa.gov/W3Browse/fermi/fermigbrst.html}.}. The extracted values are only accepted if their ratio falls into the observed distribution of $F_{\gamma}/T_{90}$.
Such a selection was performed in order to ensure that the simulated GRB sample only contains physical sources, i.e. bursts with values of the ratio $F_{\gamma}/T_{90}$ compatible with observed GRBs, given that this parameter is a key in the determination of the isotropic gamma-ray luminosity in the observer frame, $L_{\gamma,\mathrm{iso}}=4 \pi d_{L}^2(z)\frac{F_{\gamma}}{T_{90}}$. Note however that fixing the value of $F_{\gamma}/T_{90}$ actually implies that each GRB of the sample is degenerate in $L_{\gamma,\rm iso}$ and $z$ (through the luminosity distance), since different combinations of these quantities might yield the same value of $F_{\gamma}/T_{90}$.
\begin{table*}
\begin{ruledtabular}
\begin{tabular}{lcccccccc}
 &\multicolumn{2}{c}{$\Gamma=100$}
&&
 \multicolumn{2}{c}{$\Gamma=300$}
&&
\multicolumn{2}{c}{$\Gamma=600$} \\\cline{2-3}\cline{5-6}\cline{8-9}
Detector & $\rm E^{*}_{\nu_{\mu},\rm max}$ [GeV] & $\theta^{*}$ [deg] && $\rm E^{*}_{\nu_{\mu},\rm max}$  [GeV] & $\theta^{*}$ [deg] && $\rm E^{*}_{\nu_{\mu},\rm max}$ [GeV] & $\theta^{*}$ [deg]     \\
\hline
KM3NeT/ORCA & 27 & 26 && 73 & 13 && 121 & 9\\
KM3NeT/ARCA & - & - && 129 & 9 && 227 & 6\\
DeepCore & 27 & 26 && 78 & 13 && 165 & 8\\
IceCube & - & - && 156 & 8 && 258 & 5\\
\end{tabular}
\caption{Energy values $\rm E^{*}_{\nu_{\mu},\rm max}$, in GeV, at which each detector is expected to observe the highest number of $\nu_{\mu}+\bar{\nu}_{\mu}$ induced events, and corresponding plane angle values $\theta^{*}$, in degrees, adopted to search for such events around GRBs, corresponding to a solid angle $\Omega=2\pi(1-\cos(\theta^{*}/2))$.}
\label{tab:bkg_angles}
\end{ruledtabular}
\end{table*}
This method was applied to SGRBs and LGRBs, separately, as to correctly characterize the two different populations.
To obtain the muon neutrino spectrum characteristic of each GRB of the sample, we considered as reference model the neutrino production simulation presented in \cite{murase}. This refers to a high-luminous GRB with bolometric (i.e. E$_1=1$~keV and E$_2=10$~MeV) isotropic gamma-ray energy $E_{\gamma,\rm iso}=10^{53.5}$ erg at $z=0.1$, resulting in a gamma-ray bolometric fluence $F_{\gamma}^* \sim 10^{-2}$ erg cm$^{-2}$.
Since the fluence values assigned to each GRB of the synthetic sample are extracted from the \textit{Fermi}-GBM catalog, that provides measurements from $e_1=10$~keV to $e_2=1000$~keV, we need to first estimate the corresponding gamma-ray fluence in the bolometric range $F_{\gamma,\rm bol}$. This correction is also known as $k$-correction \cite{bloom}:
\begin{equation}
  k = \frac{F_{\gamma,\rm bol}}{F_\gamma} = \frac{F_{\gamma}\rm\left [ \frac{E_1}{1+z}, \frac{E_2}{1+z} \right ]}{F_{\gamma}\left [e_1,e_2 \right ]},
\end{equation}
Given the linear scaling among neutrino and gamma-ray fluences predicted by the collisionally heating GRB model, we can rescale the neutrino fluence predictions from \cite{murase} by a factor $\sim F_{\gamma,\rm bol}/F^*_\gamma$. 
This factor could be obtained directly from gamma-ray spectra specifically for each GRB. However, since the shape of the gamma-ray spectrum does not enter additionally into our computations, and since most of GRB spectra can be described by the same functional form (the Band function discussed in Sec.~\ref{sec:introduction}), we decided to include a median correction in our analysis. In order to do so, we collected all the $k$-corrections calculated based on the GRB sample observed by \textit{Fermi}-GBM during the first ten years of its operation and with known redshift ($\sim$4.5\% out of the total sample) \cite{k_corrections} and computed the median value of such a distribution. Then, each time a value of gamma-ray fluence was extracted in a synthetic GRB of the sample, this was $k$-corrected with the median of such distribution, i.e. $\bar{k}=1.13$ (see Fig.~\ref{fig:k_corr}):
\begin{equation}
    F_{\gamma,\rm bol}=\bar{k} \cdot F_{\gamma} = 1.13 \cdot F_{\gamma}.
\end{equation}
In principle, different $k$-corrections should be applied to the two samples of LGRBs and SGRBs, because these manifestly show different spectral slopes of the emitted prompt radiation. However, the \textit{Fermi}-GBM sample from which the $k$-correction is extracted \cite{k_corrections} is quite limited: it contains only sources with known redshift (amouting to 13 SGRBs and 122 LGRBs). As a result, the sample of SGRBs for which the $k$-correction has been evaluated is not statistically relevant to derive a physically motivated $k$-correction different than that of LGRBs. Hence, we will adopt the median value for both populations. \\
In the data sample of up-going events collected by a neutrino telescope, the background is mainly constituted by atmospheric neutrinos \cite{gaisser}. The Honda model is adopted as a reference for the atmospheric neutrino flux \cite{honda}. The number of events expected in coincidence with the burst depends on: i) its duration, namely the temporal window defined by $T_{90}$; ii) the search angular window around the GRB position, i.e. the solid angle $\Omega=2\pi(1-\cos(\theta/2))$, where $\theta$ is the aperture of the search cone. The aperture of the search cone should be carefully chosen as to maximize the detection prospects of a signal against the background. Given a model, determined by a specific value of $\Gamma$, we adopted the same aperture cone angle for all the simulated GRBs, the angle changing with the value of $\Gamma$. We conservatively set it to $\theta^{*}=3\theta_{\nu \mu}(\rm E^{*}_{\nu_{\mu},\rm max})$, where $\theta_{\nu \mu}(\rm E_{\nu})=0.7^{\circ}/(\rm E_{\nu}[\rm TeV])^{0.7}$\cite{review2000} is the kinematic angle between the incoming neutrino and the emerging muon directions, while $\rm E^{*}_{\nu_{\mu},\rm max}$ is the energy at which each neutrino telescope would observe the maximum number of neutrinos expected according to the model (which depends on $\Gamma$). To obtain this value, we convolved the energy-dependent effective area $\rm A_{eff}(E_{\nu_{\mu}})$ of the detectors with the expected differential energy fluence $\rm \frac{dN(E_{\nu_{\mu}})}{dE_{\nu_{\mu}} dS}$ predicted by the model, obtaining the so-called \emph{parent function}, and then we multiplied it by the width of each energy bin $\Delta E_{\nu_{\mu}}$, as follows:
\begin{equation}
    \rm N_{\nu_{\mu}}(E_{\nu_{\mu}})= A_{eff} (E_{\nu_{\mu}}) \left(\frac{dN(E_{\nu_{\mu}})}{dE_{\nu_{\mu}} dS}\right) \Delta E_{\nu_{\mu}}.
\end{equation}
Then, for each distribution $\rm N_{\nu_{\mu}}(E_{\nu_{\mu}})$ we looked for the energy $\rm E^{*}_{\nu_{\mu},\rm max}$ corresponding to the maximum number of expected events for each $\Gamma$ and each detector.
The values so determined are given in Tab.~\ref{tab:bkg_angles}, together with the corresponding opening angle of the angular window.
For the low-energy neutrino telescopes, namely KM3NeT/ORCA and DeepCore, the solid angle opened around each GRB changes from $\Omega\simeq0.2~\rm sr$ to $\Omega\simeq0.02~\rm sr$ for Lorentz factor values from $\Gamma=100$ to $\Gamma=600$, respectively. Indeed, according to the model, the greater the value of $\Gamma$ and the higher the mean energy of the emitted neutrinos (see Eq.~\eqref{eq:peak_energy}), hence the higher the energy peak of the detectable neutrino sample. 
For values of $\Gamma\geq300$, where the energy peak in the neutrino spectrum is expected beyond 100 GeV, joint analyses that include both low and high-energy neutrino detectors (namely KM3NeT/ARCA and IceCube) are possible.
In the search performed with high-energy neutrino telescopes, the solid angle where background evaluation is performed is rather within the range $\Omega\simeq0.02~\rm sr$ and $\Omega\simeq0.01~\rm sr$. 
In the combined investigations, different angular windows are set for the different detectors in order to optimize the search, as given in Tab.~\ref{tab:bkg_angles}. For further details on the choice of the aperture cone we defer the reading to Appendix \ref{appA}.
In the following, we will evaluate the perspectives of a search over a sample comprehensive of neutrino interactions in low (DeepCore and KM3NeT/ORCA) and high (IceCube and KM3NeT/ARCA) energy specialized detectors.

\begin{figure*}[t]
\subfigure[\label{fig:neutrino_spectra}]{\includegraphics[width=\columnwidth]{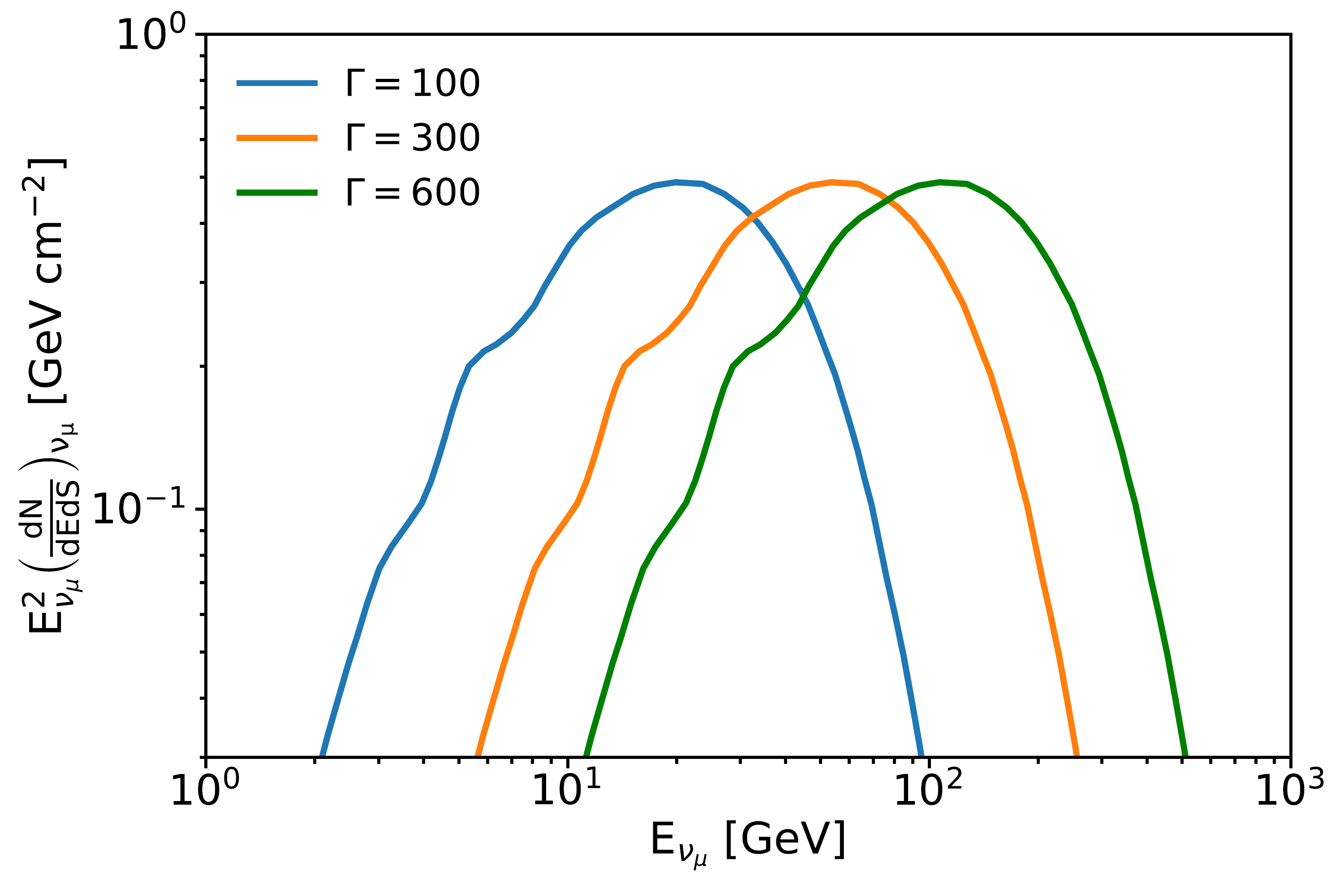}}
\subfigure[\label{fig:neutrino_ns}]{\includegraphics[width=\columnwidth]{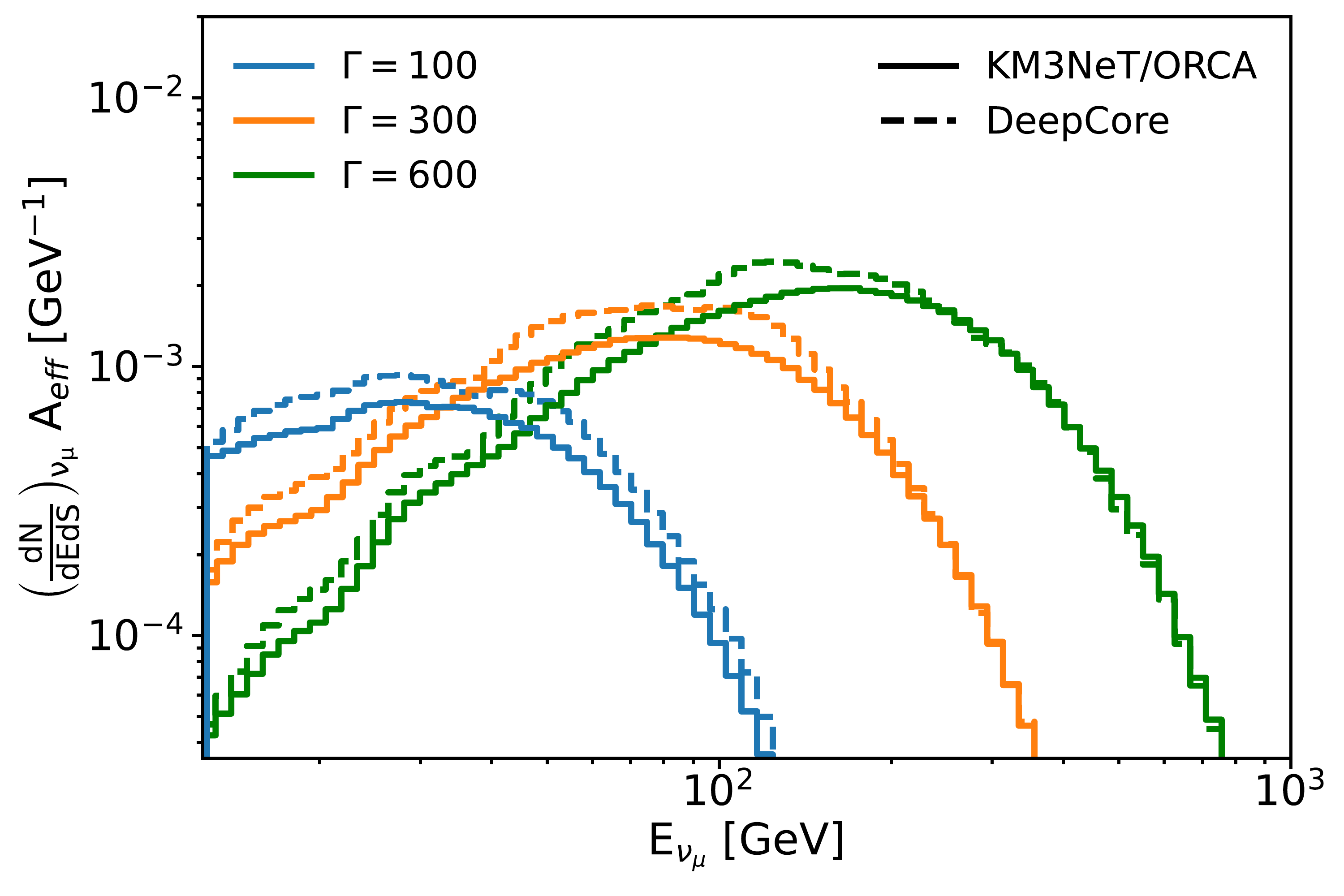}}
    \caption{(a) The $\nu_{\mu}+\bar{\nu}_{\mu}$ energy fluence for a GRB with observed gamma-ray fluence $F_{\gamma}\sim2\times10^{-3}$ erg cm$^{-2}$ for $\Gamma$=100 (cyan), $\Gamma$=300 (orange) and $\Gamma$=600 (green). (b) Number of signal events per GeV expected in KM3NeT/ORCA (solid lines) and DeepCore (dotted lines) for the neutrino energy fluences shown in (a). 
    These results are all given at trigger level.}
    \label{fig:fig_model}
\end{figure*}
\begin{table*}[t]
\centering
\begin{tabular}{lccc} 
\hline
\hline
 &\multicolumn{3}{c}{$n_s$}
 \\\cline{2-4}
Detector & $\Gamma=100$ & $\Gamma=300$ & $\Gamma=600$ \\
\hline
KM3NeT/ORCA & 4$\times$10$^{-2}$ & 7$\times$10$^{-2}$ & 1$\times$10$^{-1}$\\
KM3NeT/ORCA+KM3NeT/ARCA & - & 9$\times$10$^{-2}$ & 2$\times$10$^{-1}$\\
DeepCore & 5$\times$10$^{-2}$ & 9$\times$10$^{-2}$ & 1$\times$10$^{-1}$ \\
DeepCore+IceCube & - & 3$\times$10$^{-1}$ & 8$\times$10$^{-1}$\\
\hline
\hline
\end{tabular}
\caption{Number of events from $\nu_{\mu}+\bar{\nu}_{\mu}$ interactions expected from a GRB with gamma-ray fluence $F_{\gamma}\sim 2\times 10^{-3}$~erg~cm$^{-2}$ in low-energy detectors (KM3NeT/ORCA and DeepCore) alone, or in a combined search with high-energy detectors (KM3NeT/ARCA and IceCube, respectively). These results are all given at trigger level.}
\label{tab:number_events}
\end{table*}

\section{Prospects for neutrino detections}
\label{sec:prospects_neutrino_detections}
We here estimate whether it would be promising to look for GRB low-energy neutrino emissions with existing and under construction telescopes. Two possibilities are explored in the following: i) the search for neutrinos from individual GRBs is discussed in Sec.~\ref{sec:prospects_1grb}, with reference to an extremely bright GRB, that would represent the most optimistic scenario for individual detection of low-energy neutrinos; ii) the quasi-diffuse neutrino search from a population of GRBs with a stacking technique implemented over approximately 5 years of data acquisition is presented in Sec.~\ref{sec:stacking_prospects}.

\subsection{Detection prospects from an individual extreme GRB} \label{sec:prospects_1grb}
For the purposes of exploring the maximum discovery potential of the collisional heating mechanism in terms of individual neutrino emissions, we here consider the case of a very fluentic GRB, namely with $F_{\gamma}\sim 2\times 10^{-3}$ erg cm$^{-2}$. This value is comparable to that of the highest fluence GRB present in the \textit{Fermi}-GBM catalogue, i.e. GRB130427A. The expected differential neutrino fluence depends on the value of Lorentz factor, particularly because of the maximum energy (see Eq.~\eqref{eq:peak_energy}): in Figure \ref{fig:neutrino_spectra} we show this dependence for different values of $\Gamma$, i.e. $\Gamma=100$, $\Gamma=300$ and $\Gamma=600$. For comparison with predictions from the internal shock model, refer to Fig.~1 of \cite{bright_grb_antares}, where the neutrino fluence expectations for GRB130427A are shown.

In order to compute the expected number of signal events in each detector, we proceeded by evaluating the parent function for each of them, as shown in Fig.~\ref{fig:neutrino_ns} for DeepCore and KM3NeT/ORCA. By integrating it over the energy, we derived the results given in Tab.~\ref{tab:number_events} concerning the number of muon tracks induced by neutrino interactions. 
Though the GRB considered in this example is characterized by a high fluence (comparable to the highest fluence ever observed among all GRBs in the \textit{Fermi}-GBM catalogue), the number of signal events observable in each neutrino telescope is quite limited ($n_{\rm s}<1$), even when the events measured with low-energy detectors are integrated with those collected by high-energy ones. Nonetheless, we suggest to implement a combined search among low and high-energy neutrino detectors, i.e. KM3NeT/ORCA+KM3NeT/ARCA and DeepCore+IceCube, in order to increase the expected signal event rate. Only in the case with $\Gamma=100$ the high-energy detectors can not provide a significant contribution, given that the expected signal is entirely below 100~GeV. According to our calculations, in order to measure $n_{\rm s}\geq1$ from a single GRB in at least one of the considered detectors, this source should be characterized by an extreme fluence, as $F_{\gamma}\geq 10^{-2}$ erg cm$^{-2}$. So far, no such kind of GRBs has ever been observed, but we cannot exclude in the future a serendipitous occurrence of a nearby and very energetic explosion.

\subsection{Stacking detection prospects} \label{sec:stacking_prospects}
The signal detection rate can be greatly increased when summing up the contribution of many GRBs. However, the same holds for the background, such that stacking techniques are necessary in order to obtain a significant detection level. In fact, selecting events restricted in an angular cone around sources allows the signal to stand out with respect to the background. In our case, since the cone aperture around each GRB is not optimized individually, a source ordering in cumulative detection significance $\sigma_{\rm tot}$ is required when sources are summed up. Thanks to the stacking procedure, the cumulative significance results higher since Poissonian fluctuations of the cumulative background are smaller than the sum of background fluctuations expected from a single source. Additionally, our stacking techniques profit of the transient nature of the sources under investigation. We here present a study on expected performances of current and under construction neutrino detectors after $\sim$5 years of stacking analysis in a half-sky search for tracks (namely only upgoing muon neutrinos).

We started by building two synthetic populations of GRBs detectable by current gamma-ray satellites in 5 years of operation, one for long and one for short GRBs, each reproducing the observed rate per year in the half of the sky equal to N$_{\rm SGRB}=75$ yr$^{-1}$ and N$_{\rm LGRB}=175$ yr$^{-1}$ \footnote{Such values were obtained from the Gamma-Ray Bursts Interplanetary Network (IPNGRB) database \url{https://heasarc.gsfc.nasa.gov/w3browse/all/ipngrb.html}.}. Each GRB of the sample is described by values of $F_{\gamma}$ and $T_{90}$, randomly extracted from \textit{Fermi}-GBM observed distributions and selected as explained in Sec.~\ref{sec:performances}. For each extracted source, we estimated the expected neutrino fluence for three different values of Lorentz factor, i.e. $\Gamma= [100,300,600]$. The average source in the sample is way less fluentic than the one considered in the previous section, being characterized by a median value of $F_\gamma \sim 8 \times 10^{-6}$~erg~cm$^{-2}$ for LGRBs and $F_\gamma \sim 6 \times 10^{-7}$~erg~cm$^{-2}$ for SGRBs. The linear scaling among gamma-ray and neutrino fluence predicted by the inelastic collisional model implies a peak value in the neutrino spectrum at the level of $2\times 10^{-3}$~GeV~cm$^{-2}$ and $1\times 10^{-4}$~GeV~cm$^{-2}$ respectively, to be compared with what shown in Fig.~\ref{fig:neutrino_spectra} for an extremely fluentic GRB. The stacking approach appears therefore to be a necessary condition to test the model. 

The expected background occurring inside the detector in coincidence with the signal from each GRB is estimated in a temporal window as wide as $T_{90}\pm 0.3T_{90}$, and in an angular window defined by values reported in Tab.~\ref{tab:bkg_angles}. Once the GRB sample was populated by $\rm N$ objects, we proceeded in the following way: (i) we selected the GRB with the highest level of significance, defined as $\sigma=n_{\rm s}/\sqrt{n_{\rm b}}$; (ii) starting from such a GRB, we added one by one the others, choosing each time the GRB that provides the maximum increase of the total level of significance $\sigma_{\rm tot}$, defined as:
\begin{equation}
\sigma_{\rm tot} (\mathrm{N})=\frac{n_{\rm s,tot}}{\sqrt{n_{\rm b,tot}}}=\frac{\sum_{i=1}^{\mathrm{N}} n_{\rm s,i}}{\sqrt{ \sum_{i=1}^{\mathrm{N}} n_{\rm b,i}}} \, .
\end{equation}
We repeated the full procedure 1000 times, obtaining the median value of significance after stacking 875 LGRBs or 375 SGRBs (acquired in $\sim$5 years of half-sky gamma-ray observations), with an uncertainty band calculated through percentiles at one and two standard deviations.

By requiring $\sigma_{\rm tot}>3$ and $n_{\rm s,tot}\geq$1 as minimum conditions to define a detection, we conclude that: \\ 
(i) a detection of subphotospheric neutrinos is possible if LGRBs are included in the search, since SGRBs alone would not provide signal enough to satisfy the requirements above, in spite of the lower background level expected with respect to LGRBs; \\
(ii) the model with $\Gamma=100$ is characterized by a median value of $\sigma_{\rm tot}$ lower than unity; \\
(iii) higher values of $\Gamma$ result into a higher probability of detecting multi-GeV neutrinos in 5 years of observation with KM3NeT/ORCA and DeepCore; \\
(iv) such a possibility is increased if high-energy detectors are integrated in the search. 

\begin{figure*}[h!]
\subfigure[\label{fig:a}]{\includegraphics[width=\columnwidth]{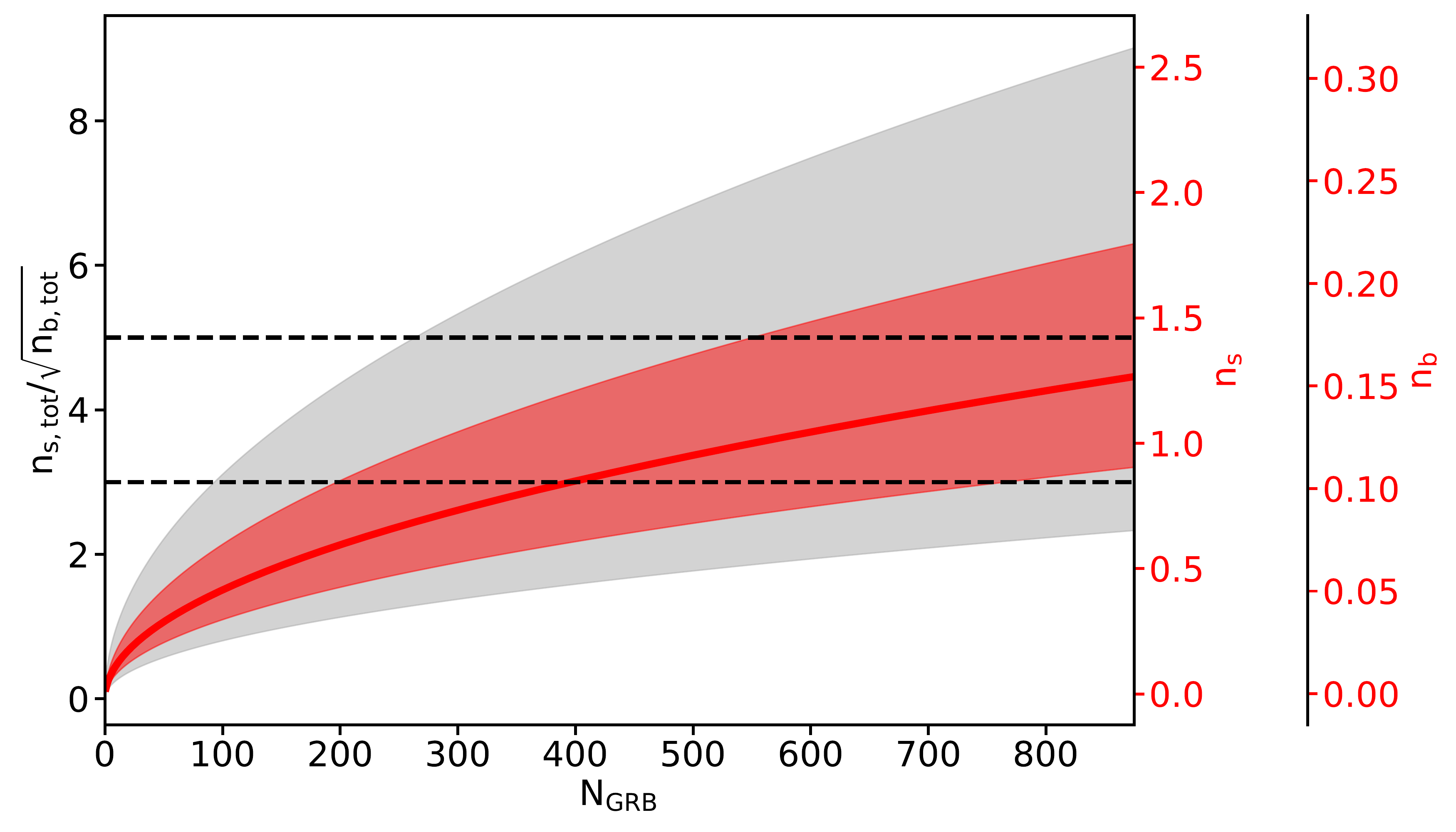}}
\subfigure[\label{fig:b}]{\includegraphics[width=\columnwidth]{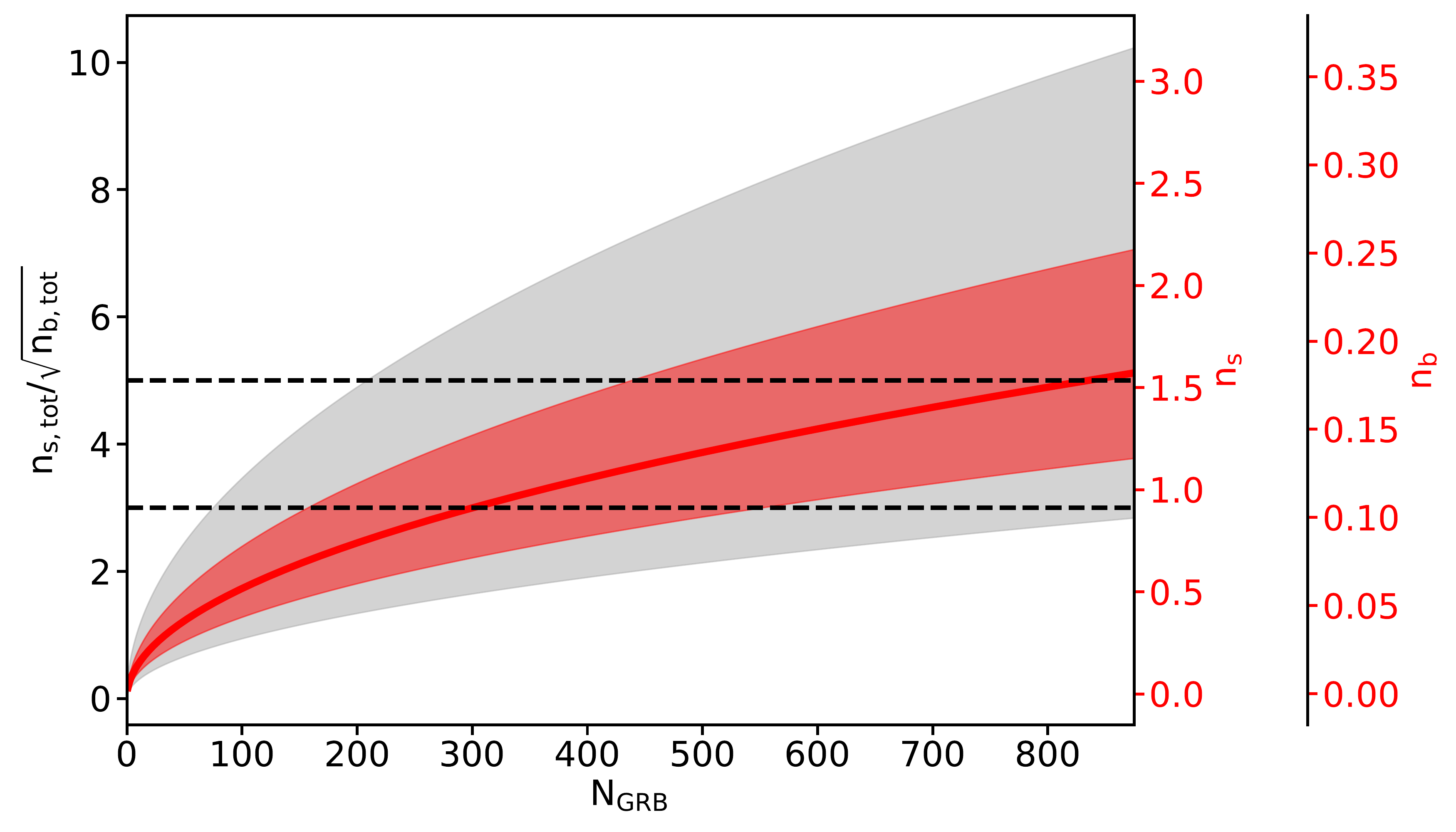}}
\subfigure[\label{fig:c}]{\includegraphics[width=\columnwidth]{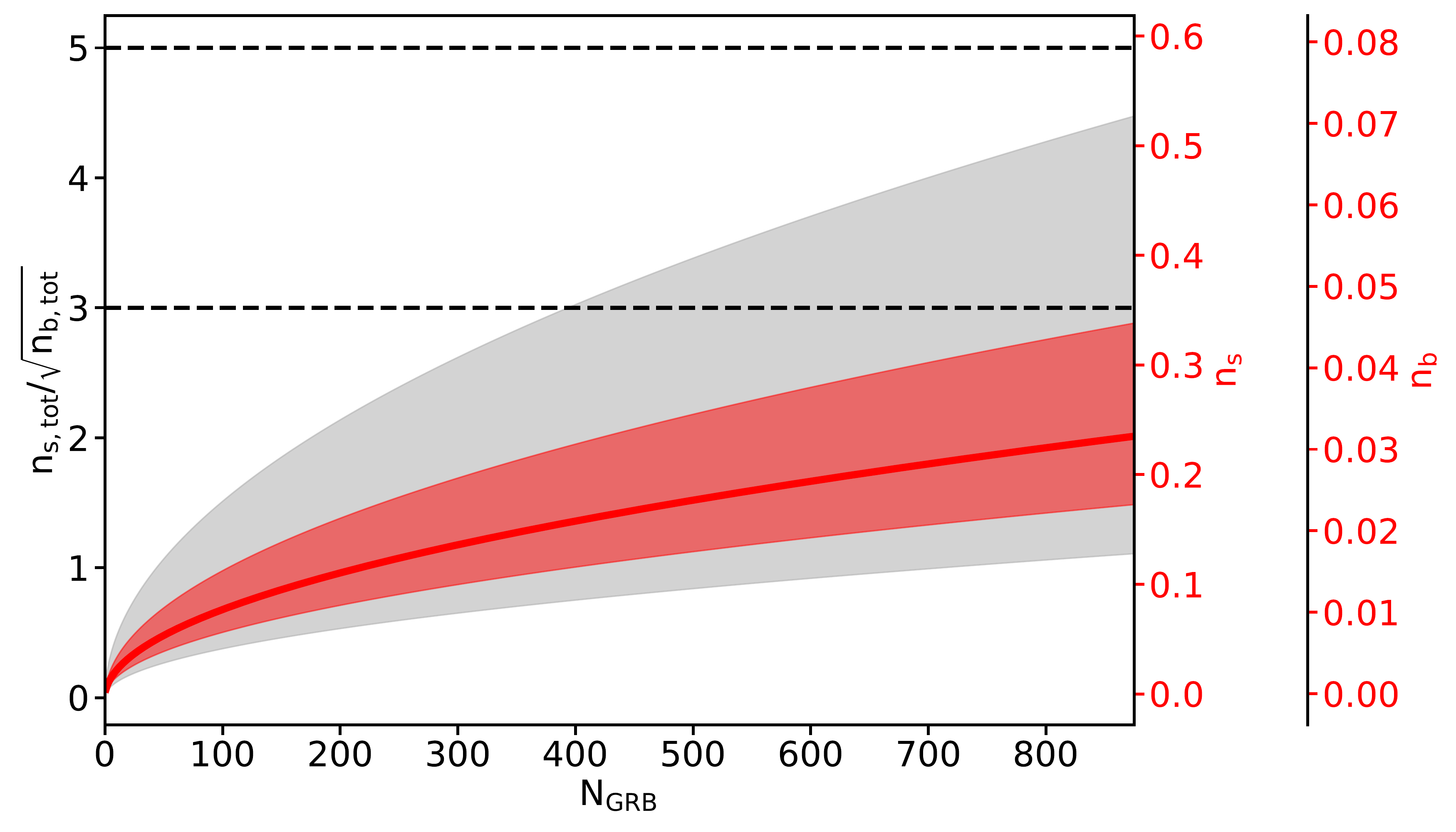}}
\subfigure[\label{fig:d}]{\includegraphics[width=\columnwidth]{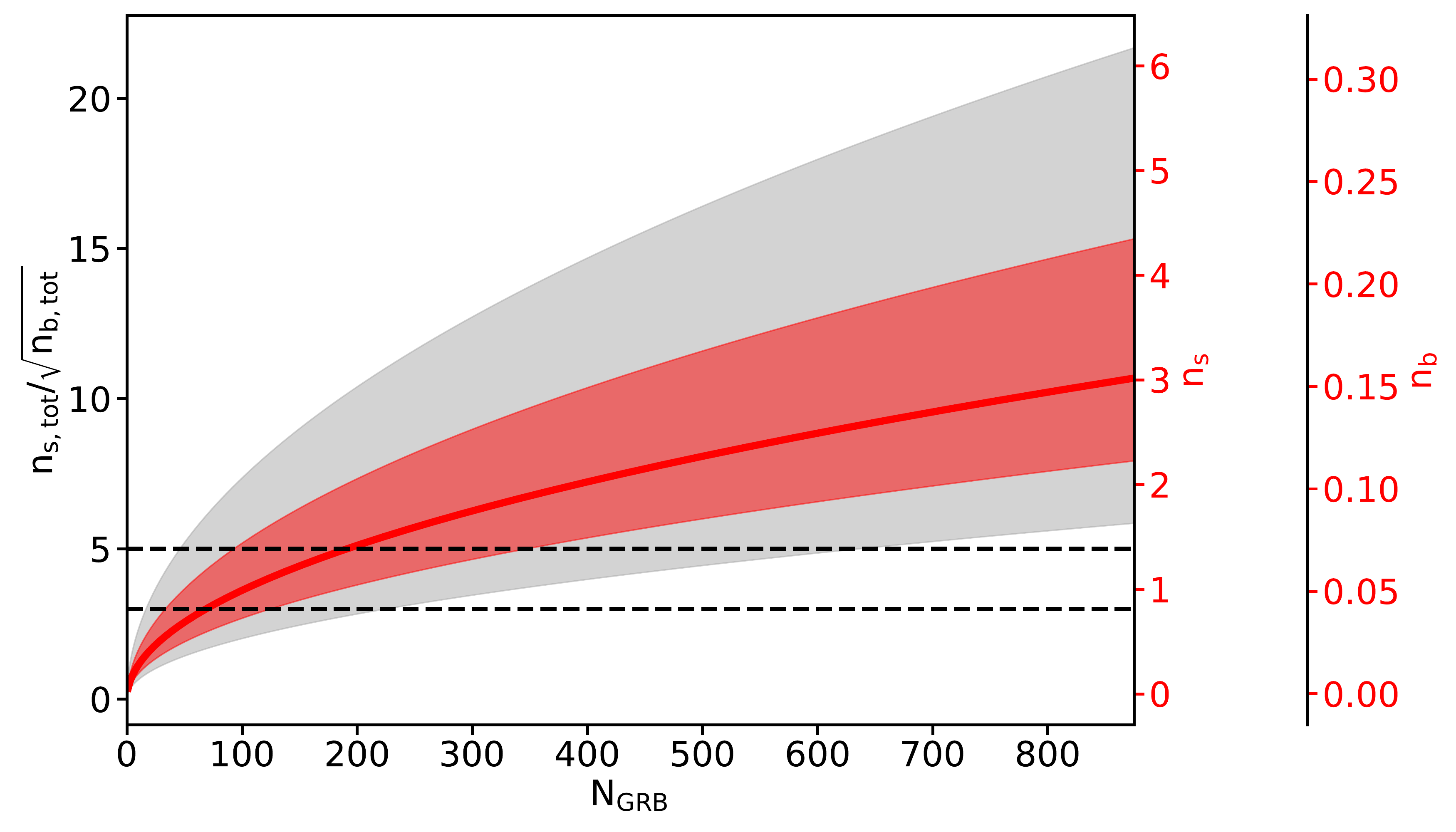}}
\subfigure[\label{fig:e}]{\includegraphics[width=\columnwidth]{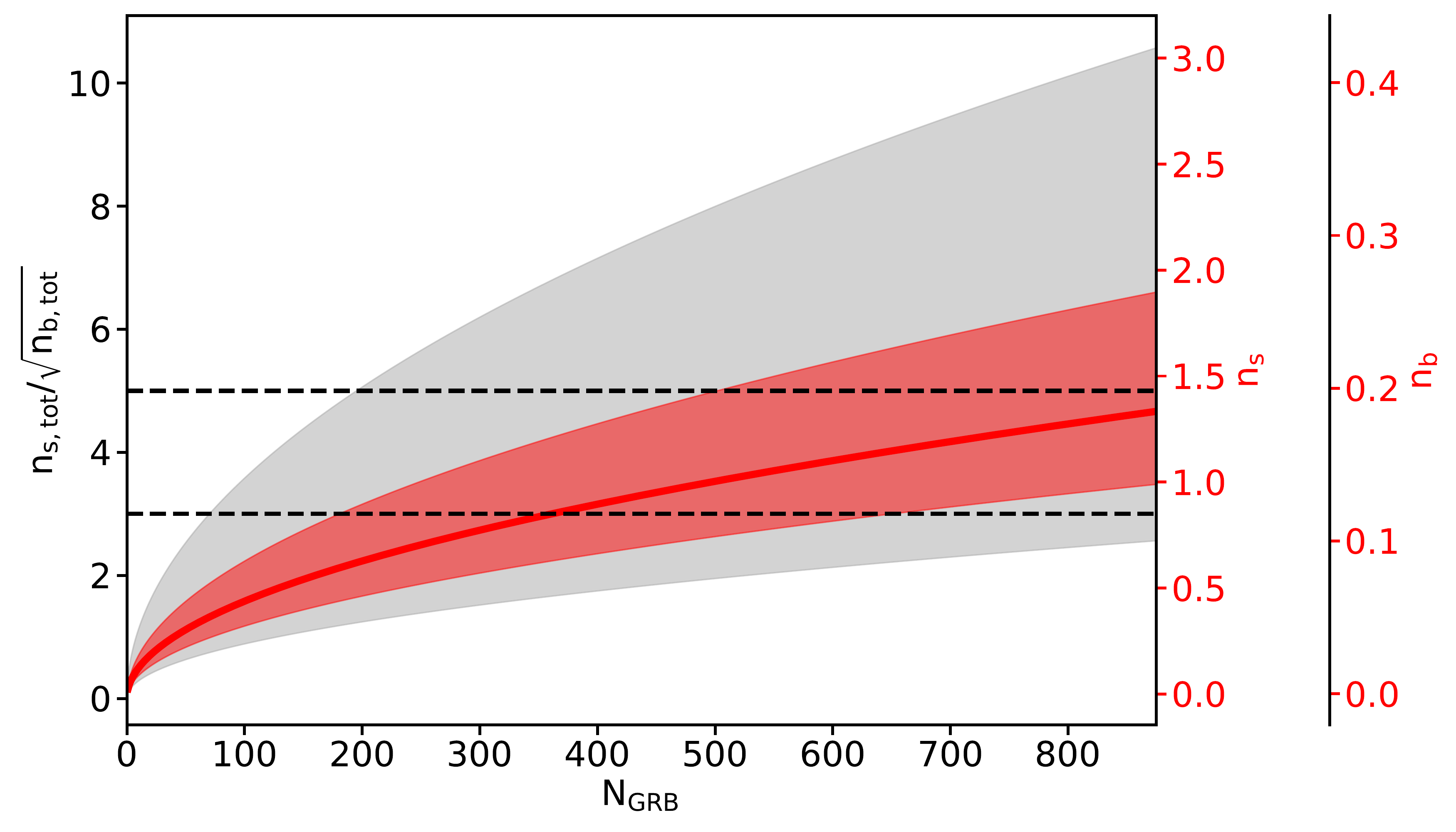}}
\subfigure[\label{fig:f}]{\includegraphics[width=\columnwidth]{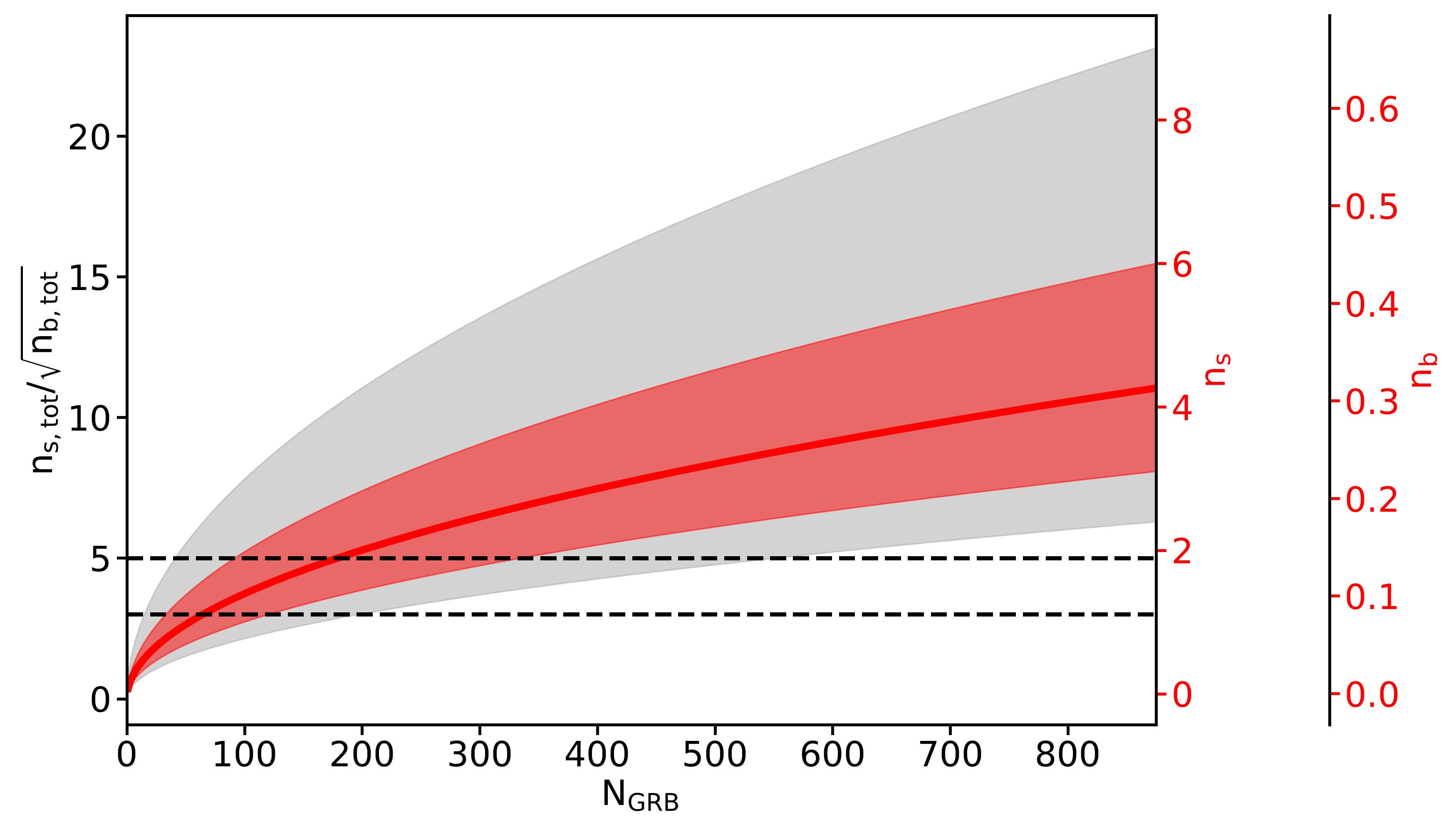}}
\caption{Level of significance $n_{\rm s,tot}/\sqrt{n_{\rm b,tot}}$ (left y-axis) achieved by stacking 875 LGRBs (equivalent to $\sim$ 5 years of half-sky search) with $\Gamma=300$, under the assumption that the gamma-ray prompt emission is originated by $pn$ collisions at sub-photospheric radii (inelastic collisional model). The level of signal and background in each detector (indicated in the right y-axis for the median result) are estimated at trigger level. Results are shown for the following neutrino detectors: (a) KM3NeT/ORCA, (b) DeepCore, (c) KM3NeT/ARCA, (d) IceCube, (e) KM3NeT/ORCA+KM3NeT/ARCA, (f) DeepCore+IceCube. The shaded red and grey regions indicate the uncertainty bands at one and two standard deviations, respectively, obtained with percentiles. The horizontal dashed black lines highlight the levels of significance $n_{\rm s,tot}/\sqrt{n_{\rm b,tot}}$=3 and $n_{\rm s,tot}/\sqrt{n_{\rm b,tot}}$=5.}
\label{fig:results_stacking}
\end{figure*}

The results obtained with $\Gamma=300$ for all detectors are shown in Fig.~\ref{fig:results_stacking}. Though a higher statistical significance can in principle be obtained by assuming $\Gamma=600$ (lower amount of background  entering the search, mostly because of the smaller angular search window selected, see Tab.~\ref{tab:bkg_angles}), we here present only the results for $\Gamma=300$, since this value is expected to more realistically describe the entire population of GRBs (see e.g. \cite{bulk_lorentz_factor}). With such a value, we obtain that there is a good chance to significantly detect multi-GeV neutrinos by stacking $\sim$900 LGRBs under the hypothesis that their prompt gamma-ray emission is explained by the collisional heating model. As visible, the overall detection significance is characterized by an extended uncertainty band, which is mostly due to the high range of values allowed for the intrinsic properties of the GRB population, namely $T_{90}$ and $F_\gamma$. While reflecting the observed distributions, the spread in GRB luminosity can lead to quite different sample realizations, which might impact the analysis results into a non significant outcome. The situation improves when combining low and high-energy detectors, as shown by the case with $\Gamma=300$ in Figs.~\ref{fig:e} and \ref{fig:f}, such that a sensitivity above $3\sigma$ is generally expected to be achieved after analyzing few hundreds of GRBs.\\
Note that the stacking procedure here implemented, once applied within the framework of the classical internal shock scenario of the fireball model, provides results in terms of expected signal neutrinos and analysis significance compatible to the limits set by the IceCube and ANTARES collaborations with respect to such a model \cite{icecube_analysis,my_analysis}. Though minor differences arise because of the adoption of trigger level effective areas and full efficiency for neutrino detectors in this work, the comparison highlights the validity of the methods here developed and the consequent conclusions.
In view of the promising results here obtained, we strongly encourage to perform optimized stacking analyses for testing the inelastic collisional model, and either confirm its occurrence or constrain the amount of GRBs possibly powered by this mechanism. \\
Note that the quasi-diffuse signal flux expected to arise as a result of the collisionally heating mechanism in the 875 LGRBs of our sample would result at the level of $\sim 2 \times 10^{-9}$~GeV~cm$^{-2}$~s$^{-1}$~sr$^{-1}$ at $\rm E_\nu \sim 100$~GeV. This value is much smaller than the atmospheric neutrino background flux at the same energy, which is rather of the order of $\sim 5 \times 10^{-4}$~GeV~cm$^{-2}$~s$^{-1}$~sr$^{-1}$. These numbers highlight the challenging task of pure diffuse searches in revealing the presence of a tiny signal, like that from the GRB population, on top of the huge atmospheric background. In turn, as Fig.~\ref{fig:results_stacking} demonstrates, the stacking approach provides a more promising perspective than the pure diffuse analysis.
In particular, the IceCube and DeepCore detectors could already have collected enough data to constrain the model here investigated and eventually confirm our finding, with the analysis technique here presented. This result encourages dedicated stacking analyses, optimized for energies lower than 1 TeV, which have not been implemented yet relatively to the GRB population at such low energies.

\section{Summary and Conclusions} 
\label{sec:conclusion}
In this work, we considered GRB jets consisting of protons and neutrons, where a fraction of the outflow kinetic energy is converted to thermal energy and radiation via inelastic nuclear collisions occurring in the photosphere. This hypothesis significantly deviates from the classical model explaining the GRB observed radiation through particles accelerated at internal shock in the optically thin region of the jet. In the photospheric scenario, a thermal spectrum is released near the photosphere and this is then modified by a dissipation mechanism related to the above mentioned $pn$ collisions. Such collisionally heated GRBs could produce multi-GeV neutrinos, namely neutrinos with energies much lower than those produced in photo-hadronic mechanisms so far investigated in the context of astrophysical neutrino data analyses. Hence we tested the possibility to detect low-energy neutrinos, on top of the expected atmospheric background, with current and under construction neutrino telescopes. As a result of our investigation, we found that these detectors are able to explore the occurrence of inelastic collisions in GRB jets by means of a stacking analysis of upgoing tracks with data collected in coincidence with $\sim$900 LGRBs. 
As previous experimental studies conducted with neutrino data indicate, it is likely that such a number of LGRBs will actually be selected for analyses after more or less 10 years of data taking \cite{antares_first_grb_analysis,my_analysis,icecube_grb2016,icecube_analysis}. We remark that the search for upgoing events requires detectors in different hemispheres to probe the entire population of GRBs, as to guarantee full sky coverage.\\
According to our results, short GRBs alone do not provide enough signal. Concerning individual GRBs, we conclude that only nearby and very energetic GRBs (rare events) can produce at least one signal event in the available detectors. The key role of neutrinos in assessing the origin of the prompt radiation emerging from GRBs demonstrates the importance of dedicated searches in the multi-GeV domain, that we encourage to start, 
by combining data collected both by low and high-energy detectors (DeepCore+IceCube and KM3NeT/ORCA+KM3NeT/ARCA). A detection of such a neutrino emission would allow to establish the baryonic nature of GRB jets, although within the context of a model that does not directly involve particle acceleration. In fact, so far, the lack of a correlation among gamma-ray signals from GRBs and neutrinos did not allow to distinguish among the possible leptonic or hadronic nature of radiation from GRB jets. Furthermore, the detection of multi-GeV neutrinos in coincidence with GRBs would provide information on the occurrence of photospheric dissipation, as well as the on the jet composition.

\section*{Acknowledgmements}
We thank the anonymous referee for the helpful comments, and constructive remarks on this manuscript. The authors acknowledge the support from the Amaldi Research Center funded by the MIUR program “Dipartimento di Eccellenza” (CUP:B81I18001170001), the Sapienza School for Advanced Studies (SSAS) and the support of the Sapienza Grant No. RM120172AEF49A82.
We gratefully acknowledge Kohta Murase for the useful discussions and the material provided, which has allowed the model investigation. We also thank Dafne Guetta for the fruitful conversations on the topics treated in the present paper.

\clearpage
\appendix

\section{Angular windows around GRBs in the background evaluation}\label{appA}
In the present section, we discuss the selection of the angular window around each GRB adopted for the background estimation. In order to reduce the very abundant background from atmospheric muons, only up-going events are considered, thus the remaining background in our study is constituted by the irreducible atmospheric neutrinos flux \cite{gaisser}, as described by the Honda model \cite{honda}.

\begin{figure}[h!]
\subfigure[\label{fig:}]{\includegraphics[width=\columnwidth]{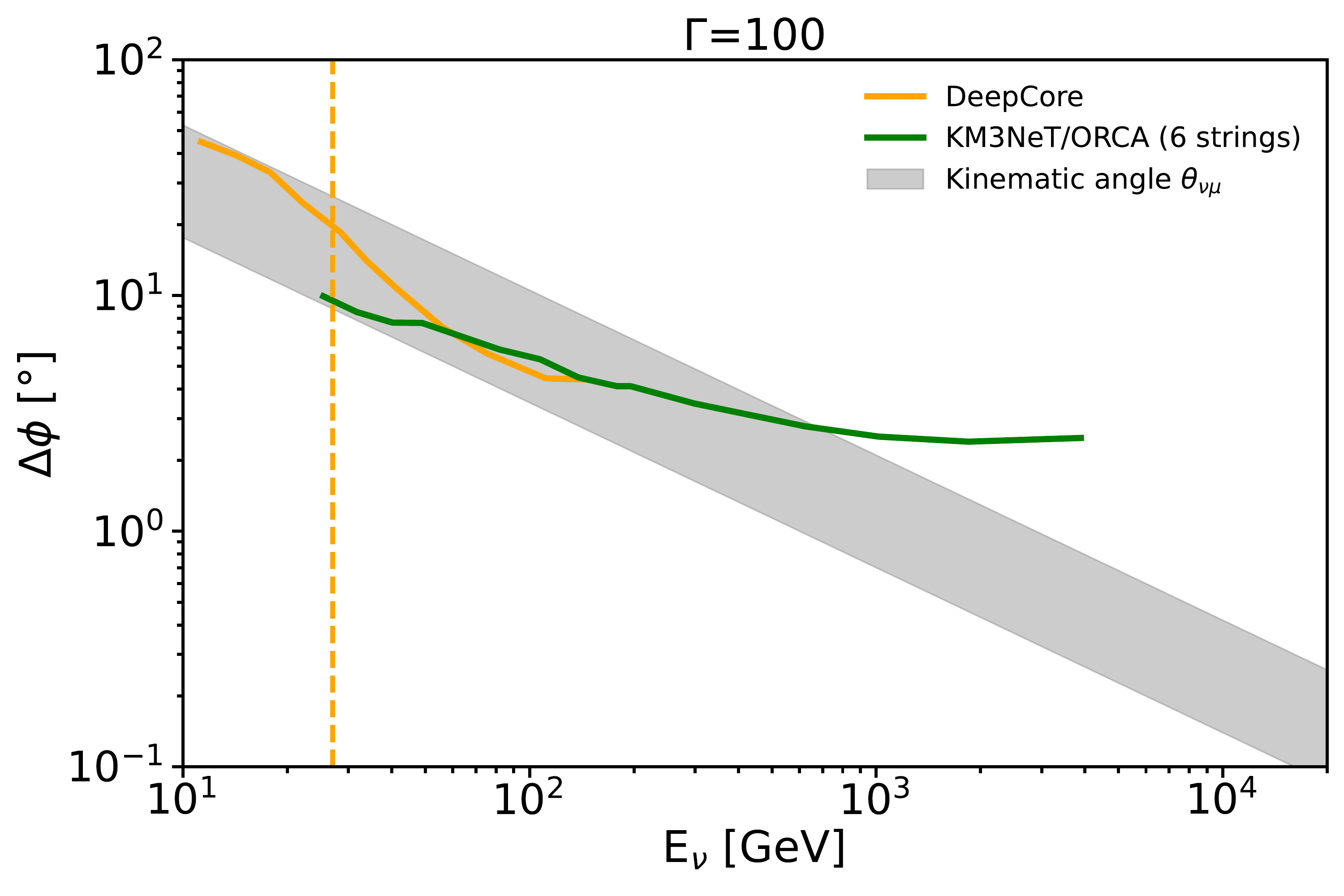}}
\subfigure[\label{fig:}]{\includegraphics[width=\columnwidth]{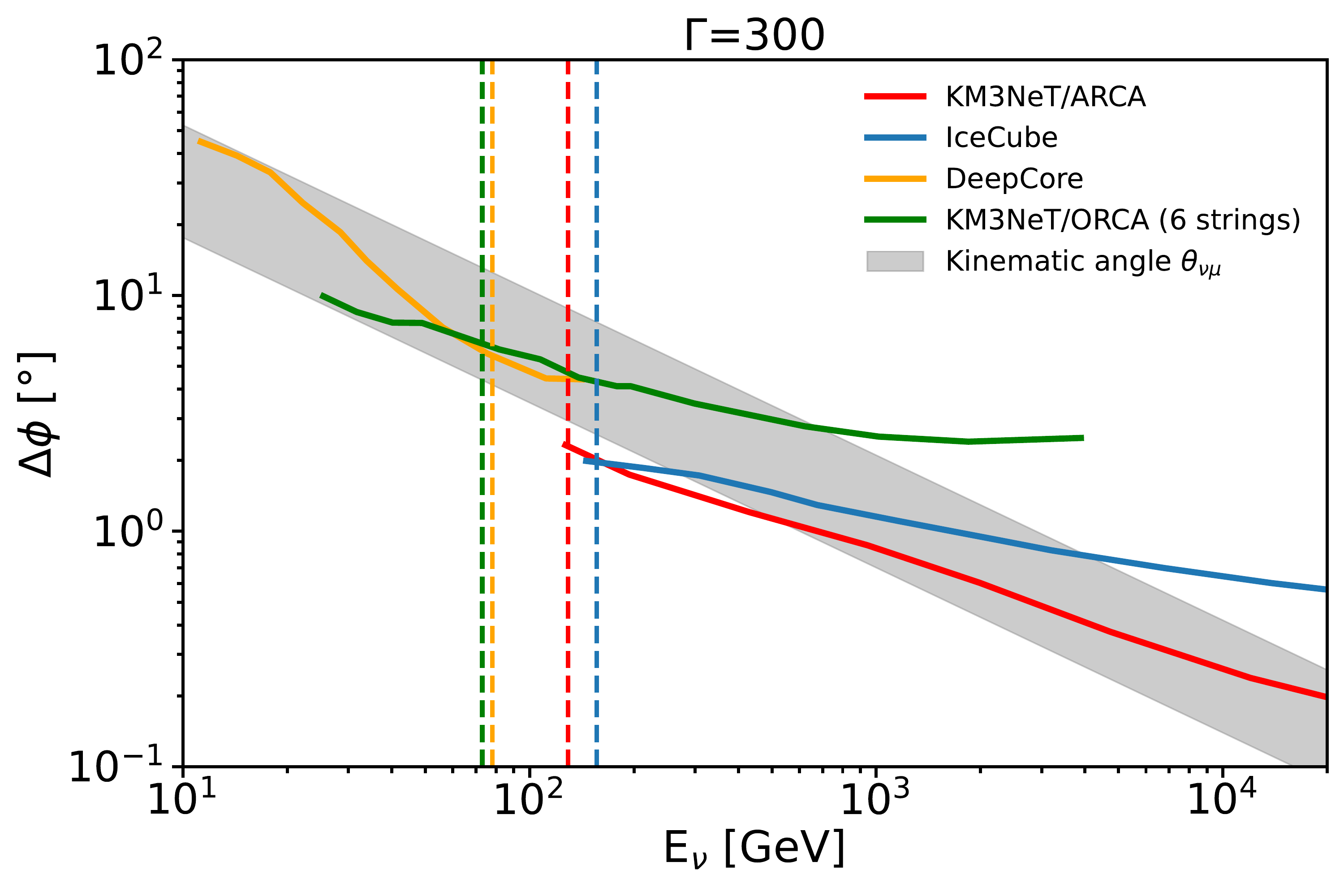}}
\subfigure[\label{fig:}]{\includegraphics[width=\columnwidth]{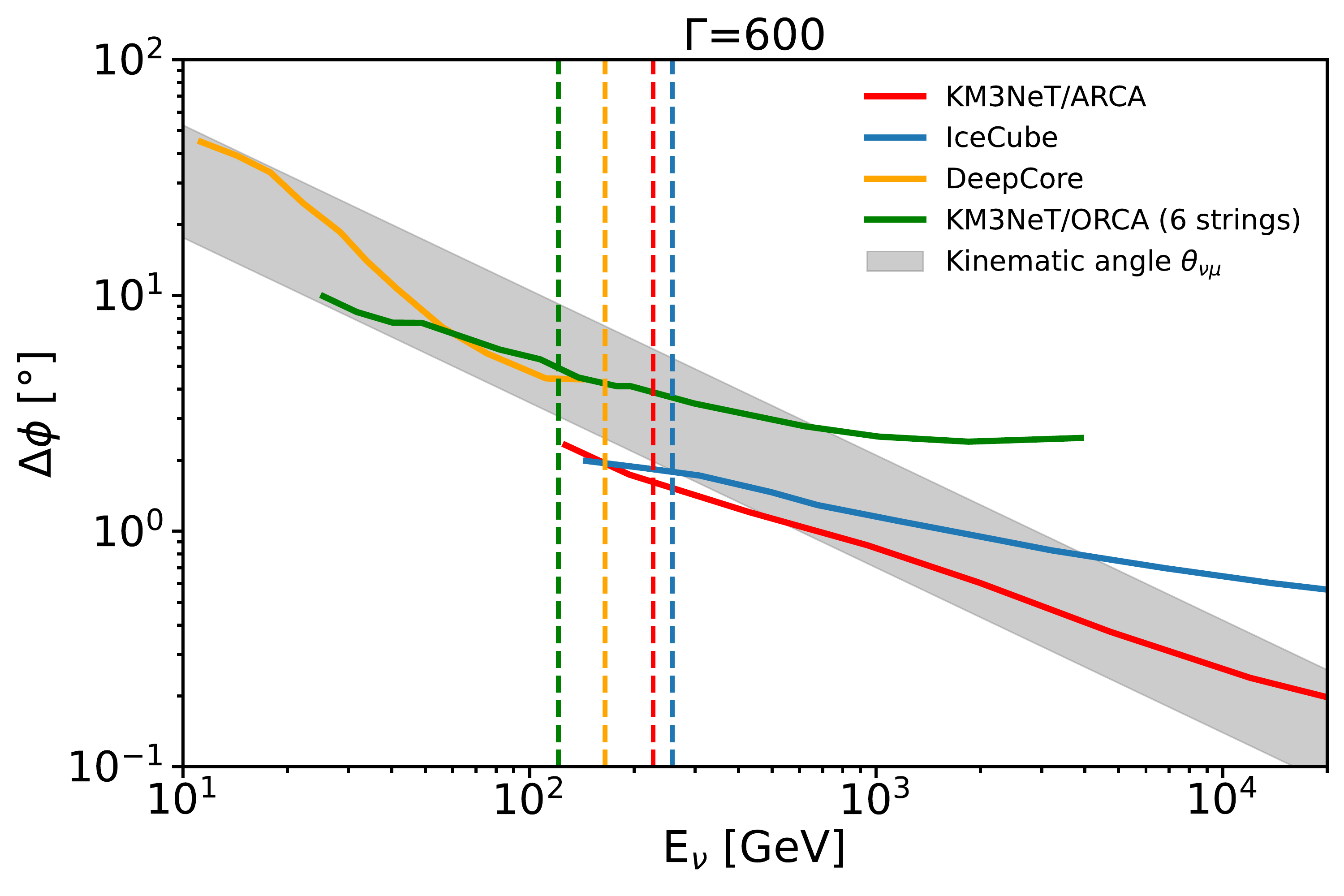}}
\caption{Median angular resolutions of $\nu_{\mu}$ charged current events for KM3NeT/ORCA (green) \cite{posICRC_feifei}, KM3NeT/ARCA (red) \cite{posICRC_ang_res_ARCA}, DeepCore (orange), and IceCube (blue) \cite{ang_res_ic_dp}. Note that the KM3NeT/ORCA angular resolution refers to a partial configuration with 6 strings, being the only available from the literature. The shaded grey region shows the interval among $\theta_{\nu{\mu}}$ and 3$\theta_{\nu{\mu}}$, being $\theta_{\nu{\mu}}$ the kinematic angle between the incoming neutrino and the muon inside the detector. The vertical dashed lines indicate the energy values $\rm E_{\nu_{\mu,max}}^{*}$ at which each detector is expected to observe the highest number of muon neutrinos events (see Tab.~\ref{tab:bkg_angles}). The three panels differ in the vertical lines, which depend on the model for the signal: in particular, (a) $\Gamma=100$, (b) $\Gamma=300$, and (c) $\Gamma=600$.}
\label{fig:ang_res}
\end{figure}

As in Cherenkov telescopes neutrinos are not detected directly, rather via the secondary particles produced in neutrino interactions (we here consider tracks of muons originated in CC $\nu_{\mu}$ interactions), two effects need to be carefully considered when inferring the original direction of the neutrino: i) the kinematic angle between the primary neutrino and the induced muon, and ii) the quality of directional reconstruction of muons. Both these factors contribute to the angular resolution of the detector. Their effects depend on the energy of the involved particles, leading to an improved angular resolution with increasing energy.
In Fig.~\ref{fig:ang_res} we show the energy-dependent kinematic angle $\theta_{\nu{\mu}}$ \cite{review2000}, in the form of a band extending from $\theta_{\nu{\mu}}$ to 3$\theta_{\nu{\mu}}$, as compared to the most updated median angular resolutions available from literature for KM3NeT/ARCA \cite{posICRC_ang_res_ARCA}, KM3NeT/ORCA \cite{posICRC_feifei}, DeepCore, and IceCube \cite{ang_res_ic_dp}. Note that the median angular resolution already includes both kinematics and detector effects. Three panels are reported in order to compare the parametrizations of the instrumental angular resolutions with the values of the kinematic angle at $\rm E_{\nu_{\mu},max}^{*}$, where each detector is expected to observe the highest number of $\nu_{\mu}$ events for different values of bulk Lorentz factor of the jet, again $\Gamma=100$, $\Gamma=300$ and $\Gamma=600$ (see Sec.~\ref{sec:performances} and Tab.~\ref{tab:bkg_angles}). We can see that, in all cases, the choice of 3$\theta_{\nu_{\mu}}$ as plane angle of the cone opened around each GRB is a conservative one, as it is larger than the median angular resolution of neutrino detectors at $E_{\nu_{\mu},max}^{*})$. For this reason, in the background estimation, we conservatively decided to open around each GRB an angular window defined by a solid angle $\Omega=2\pi(1-\cos(\theta^{*}/2))$ with $\theta^{*}=3\theta_{\nu{\mu}}(\rm E_{\nu_{\mu},max}^{*})$.

\clearpage
\bibliography{paper}

\end{document}